\documentclass[orivec]{llncs}
\usepackage{microtype}

\usepackage{amsmath}
\usepackage{amssymb}
\usepackage{bussproofs}
\usepackage{graphicx}
\usepackage{comment}
\usepackage{color}
\usepackage{textcomp}
\usepackage{fixltx2e}
\usepackage{multirow}
\usepackage{multicol}
\usepackage{rotating}
\usepackage{url}
\usepackage{enumitem}

\newcommand{\limp}[0]{\ensuremath{\rightarrow}}

\newcommand{\mimp}[0]{\ensuremath{{-\!*\;}}}

\newcommand{\simp}[0]{\ensuremath{\triangleright}}

\newcommand{\mand}[0]{\ensuremath{*}}

\newcommand{\dmd}[0]{\ensuremath{\Diamond}}

\newcommand{\lsfoasl}[0]{\ensuremath{LS_{FOASL}}}

\newcommand{\psl}[0]{\ensuremath{\mbox{PASL}}}
\newcommand{\lspasl}[0]{\ensuremath{LS_{PASL}}}

\newcommand{\lssl}[0]{\ensuremath{LS_{SL}}}
\newcommand{\foasl}[0]{\ensuremath{\mbox{FOASL}}}

\newcommand{\myseq}[3]{\ensuremath{#1; #2 \vdash #3}}

\newcommand{\pfun}{\rightharpoonup}

\newcommand{\lvar}{\rm LVar}

\def\Acal{\mathcal{A}}
\def\Rcal{\mathcal{R}}
\def\Vcal{\mathcal{V}}
\def\Ccal{\mathcal{C}}
\def\Fcal{\mathcal{F}}
\def\Gcal{\mathcal{G}}
\def\Mcal{\mathcal{M}}
\def\Ncal{\mathcal{N}}
\def\Lcal{\mathcal{L}}

\def\Rcal{\mathcal{R}}

\def\Ical{\mathcal{I}}

\def\int{\mathit{Int}}

\def\Abb{\mathbb{A}}
\def\Cbb{\mathbb{C}}
\def\Ebb{\mathbb{E}}

\def\CSbb{\mathbb{CS}}

\def\Nbb{\mathbb{N}}
\def\Pbb{\mathbb{P}}
\def\Dbb{\mathbb{D}}

\newcommand{\exfml}[1]{Ex#1}

\def\nil{\mathit{nil}}

\def\apointsto{\leadsto}
\def\fpt{\mathfrak{f}}
\newcounter{contenum}

\begin{document}







\title{Completeness for a First-order Abstract Separation Logic}

\author{Zh\'e H\'ou \and Alwen Tiu}
\institute{Nanyang Technological University, Singapore\\
\texttt{zhe.hou@ntu.edu.sg, atiu@ntu.edu.sg}}



\maketitle




\begin{abstract}
Existing work on theorem proving for the assertion language of
separation logic (SL) either focuses on abstract semantics which are not
readily availabe in most applications of program verification,
or on concrete models for which completeness is not possible.
An important element in concrete SL is the points-to predicate
which denotes a singleton heap. SL with the points-to predicate
has been shown to be non-recursively enumerable. In this paper,
we develop a first-order SL, called FOASL, with an abstracted
version of the points-to predicate. We prove that FOASL is sound
and complete with respect to an abstract semantics, of which the
standard SL semantics is an instance. 
We also show that some reasoning principles involving the 
points-to predicate can be
approximated as FOASL theories, thus allowing our logic to be used
for reasoning about concrete program verification problems. 
We give some example theories that are sound with respect to 
different variants of separation logics from the literature, including
those that are incompatible with Reynolds's semantics.
In the experiment we demonstrate our FOASL 
based theorem prover which is able to handle a large fragment 
of separation logic with heap semantics as well as non-standard semantics. 
\end{abstract}

\section{Introduction}
\label{sec:intro}

Separation Logic (SL) is widely used in program verification and
reasoning about memory models~\cite{OHearn2001,reynolds2002}. SL
extends the traditional Hoare Logic with logical connectives
$\mand,\mimp$ from the logic of Bunched Implications (BI). These new
connectives in BI provide an elegant way to reason about resources
locally, enabling analyses of large scale programs. Current 
work on SL can be divided into two categories: one on the abstract
separation logics and the other on the concrete ones. On the abstract side
there has been study on BI and its Boolean variant
BBI~\cite{Ohearn1999,Wendling2010}. Closely related are abstract
separation logic and its neighbours~\cite{calcagno2007,Wendling2009}.
Abstract separation logics lack the interpretation of the points-to
predicate $\mapsto$, which represents a single memory cell. In this
setting, the semantics is just an extension of commutative monoids
with certain properties usually called separation
theory~\cite{dockins2009,brotherston2013}. On the concrete side there
have been developments along several directions, such as proof methods for
SL with memory model semantics~\cite{galmiche2010,hou2015,Lee2013},
and symbolic
heaps~\cite{berdine2005b,berdine2005,Navarro2011,brotherston2011}. There
have been numerous modifications of SL, e.g., Fictional Separation
Logic~\cite{JensenBirkedal2012}, Rely/Guarantee~\cite{vafeiadis2007},
Concurrent Separation Logic~\cite{Brookes2007}.

To support reasoning about Hoare triples, it is essential to have 
proof methods for the assertion logic.
In the reminder of this paper we
focus on the assertion logic of separation logic. Although theorem
proving for propositional abstract separation logics ($\psl$s) is
undecidable~\cite{Wendling2010,Brotherston13}, there have been
semi-decision procedures for those
logics~\cite{Brotherston2010,Park2013,hou2013b,brotherston2013,larcheywendling2014}.
However, since $\psl$s do not impose a concrete semantic model, they
usually cannot be directly used in program verification.  
The situation is more intriguing for separation logic with
memory model semantics. Calcagno et al. showed that the full logic is
not recursively enumerable~\cite{Calcagno01},
thus it is not possible to have a sound and complete finite proof system
for the full separation logic. 
Interestingly,
their result uses only an extension of first-order logic, without function symbols,
but with a two-field points-to predicate, i.e., predicates of the form
$[a \mapsto c,d]$, which represents a memory cell with a record of two fields.
Recent study also shows that the points-to predicate $\mapsto$ is
a source of undecidability. For example, restricting the record field
(i.e., right hand side) of $\mapsto$ to one variable and allowing only
one quantified variable reduces the satisfiability of SL to
PSPACE-complete~\cite{demri2014b}.
The above work indicates that directly handling the points-to
predicate in the logic may not be easy. Section 4 of~\cite{hou2015}
details related issues on various proof systems for separation logic
with memory model.
Another complicating factor in designing a proof system for separation logic 
is the ``magic wand'' $\mimp$ connective. The separation conjunction 
$\mand$ can be encoded using $\mimp$, but not the
other way around~\cite{brochenin2012,demri2014}.  Consequently, most
proof methods for SL with concrete semantics are restricted to
fragments of SL, typically omitting $\mimp$. The connective $\mimp$,
however, has found many applications, such as
tail-recursion~\cite{maeda2011}, iterators~\cite{neelakantan2006},
``septraction'' in rely/guarantee~\cite{vafeiadis2007}, amongst other
motivating examples discussed in the introduction of~\cite{Lee2013}.

Since completeness with respect to Reynolds's model is not possible,
an interesting question arises as to what properties of points-to
($\mimp$ is not so crucial in terms of the completeness property) one
should formalize in the system, and what kind of semantics the
resulting proof system captures. There have been at least a couple of
attempts at designing a proof system that features both $\mimp$ and
points-to; one by Lee and Park~\cite{Lee2013} and the other by Hou et
al. \cite{hou2015}.  In \cite{Lee2013}, Lee and Park claimed
incorrectly that their proof system is complete with respect to
Reynolds's semantics, though this claim was later retracted.\footnote{
  See \url{http://pl.postech.ac.kr/SL/} for the revised version of
  their proof system, which is sound but not complete
  w.r.t. Reynolds's semantics.}  In \cite{hou2015}, Hou et al. give a
proof system $\lssl$ that is sound with respect to Reynolds's
semantics, but no alternative semantics was proposed nor any
completeness result stated.  It is also not clear in general what
proof rules one should include in the proof system such as that in
\cite{hou2015}. This has led to the introduction of various ad hoc
proof rules in \cite{hou2015}, which are often complex and unintuitive, 
in order to capture specific properties 
of the underlying concrete separation logic models (e.g., Reynolds's model),
resulting in a complex proof system, which is both hard to implement 
and hard to reason about. 




In this paper, we revisit our previous work~\cite{hou2015}
in order to provide an abstract semantics and 
a sound and complete proof system with respect to that abstract semantics,
that are useful for reasoning about the meta-theory of the
proof system, and easy to extend to support reasoning about various
concrete models of separation logic. 
Our point of departure is
to try to give a minimal proof system and encode as many properties of points-to as possible using
{\em logical theories}, rather than proof rules, and to formalize as
inference rules only those properties that cannot be easily encoded as
theories. This led us to keep only two proof rules for the points-to predicate
from \cite{hou2015} (see the rules $\apointsto_1$ and $\apointsto_2$ in Figure~\ref{fig:LS_FOASL} in Section~\ref{sec:lsfoasl}). 
Semantically, these two rules are justified by 
a new semantics of an abstract version of the points-to predicate (notationally
represented by $\apointsto$ here, to distinguish it from the points-to predicate $\mapsto$ in the
concrete SL), that is, 
it is a function that constructs a heap from 
a tuple of values. In particular, we do not assume that the constructed
heap from a tuple is a singleton heap, so a points-to predicate such as
$[a \apointsto b,c]$ in our semantics denotes a heap, but not necessarily
a singleton heap mapping $a$ to a two-field record $(b,c).$
Reasoning in the concrete models, such as Reynolds's SL, which 
may require properties of singleton heaps, can be approximated by
adding theories to restrict the interpretation of points-to
(see e.g., Section~\ref{sec:reynolds}). Obviously one would not be able 
to completely restrict the interpretation of $\apointsto$ to singleton heaps via
a finite theory, as one would then run into the same incompleteness
problem as shown in \cite{Calcagno01}.

The proof system for the first-order abstract separation logic that we introduce here, 
called $\lsfoasl$, is based on the proof system $\lspasl$ for propositional
abstract separation logic ($\psl$)~\cite{hou2013b}, which is a labelled sequent calculus.
We choose labelled calculi as the framework to formalize our logic
since it has been shown to be a good framework 
for proof search automation for abstract separation logics~\cite{hou2015,larcheywendling2014}. 
Formulas in a labelled sequent in $\lspasl$ are interpreted relative to the
interpretation of the labels they are attached to, so a labelled formula such as
$h : F$, where $h$ is a label and $F$ is a formula, 
denotes the truth value of $F$ when evaluated against the heap $h$.
In extending $\psl$ to the first-order case, especially
when formulating theories for specific concrete models, it turns out that
we need to be able to assert that some formulas hold universally for all heaps,
or that they hold in some unspecified heap, both of which are not expressible in $\psl$. 
To get around this limitation, we introduce modal operators to allow 
one to state properties that hold globally in all heaps or in some heap.
Section~\ref{sec:axioms} shows some
examples of uses of these modal operators in theories 
approximating concrete models of separation logics.

The semantics of $\foasl$ is formally defined in
Section~\ref{sec:foasl}. In Section~\ref{sec:lsfoasl} we present the
proof system $\lsfoasl$, which is an extension of $\lspasl$ with rules
for first-order quantifiers, the points-to predicate and modal
operators.  In Section~\ref{sec:cmc}, we prove soundness and
completeness of $\lsfoasl$ with respect to the semantics described in
Section~\ref{sec:foasl}. The completeness proof is done via a
counter-model construction similar to that
in~\cite{wendling2012,hou2013b} for the propositional case, but our
proof is significantly more complicated as we have to deal with the
first-order language and the construction of the model for the
points-to predicate, which requires a novel proof technique.  We show
in Section~\ref{sec:axioms} that all the inference rules for points-to
in the labelled sequent calculus $\lssl$~\cite{hou2015} can be derived
using theories in our logic, except one rule called $HC$, which is not
used in the current program verification tools. Our theories for
points-to cover the widely-used symbolic heap fragment of SL, thus our
logic can be used in many existing program verification
methods. Furthermore, we can also prove many formulae that are valid
in SL but which cannot be proved by most existing tools. An implementation
is discussed in Section~\ref{sec:imp}, we show that our prover can
reason about the standard heap semantics and the non-standard ones.

\section{First-order Abstract Separation Logic}
\label{sec:foasl}

This section introduces First-order Abstract Separation Logic
($\foasl$). The formulae of $\foasl$ are parameterized by a first order
signature $\Sigma = (\Rcal, \Ccal)$, consisting of a set of predicate
symbols $\Rcal$ and a set of constants $\Ccal$.  Constants are ranged
by $a$, $b$ and $c$, and predicate symbols by $p$ and $q$ (possibly
with subscripts). We also assume an infinite set $\Vcal$ of first-order
variables, ranged over by $x, y, z$.  A {\em term} is either a
constant or a variable, and is denoted by $e$, $s$ and $t$.  
We assume that $\Rcal$ includes a
symbol $=$, denoting the equality predicate, and a finite collection
of \emph{abstract points-to} predicates.
We shall use the notation $\apointsto^n$ 
to denote an abstract points-to predicate of arity $n$. We use an infix notation when
writing the abstract points-to predicates.
For an abstract points-to predicate of arity $k$, 
taking arguments $t_1,\dots,t_{k}$, we
write it in the form:
$$
t_1 \apointsto^k t_2, \ldots, t_{k}.
$$
We shall omit the superscript $k$ when the arity of $\apointsto$ is
not important or can be inferred from the context of discussion.  Note
that the abstract points-to $\apointsto$ is not the points-to
predicate in SL, but is a weaker version whose properties will be
discussed later.
To simplify presentation, we do not consider function symbols in our language,
but it is straightforward to add them. In any case, the incompleteness result
for concrete SL of \cite{Calcagno01} holds even in the absence of function symbols, 
so the omission of function symbols from our logic does not make the completeness proof
for our logic conceptually easier.

The formulae of $\foasl$ are given by the following grammar:
$$
\begin{array}{ll}
F ::= & \top^* \mid \bot \mid p(t_1,\dots,t_k) \mid s \apointsto t_1,\dots,t_l \mid s = t 
\mid F \limp F \mid\\
& F \mand F \mid F \mimp F \mid \dmd F \mid \exists x. F
\end{array}
$$
The logical constant $\bot$, the connective $\limp$
and the quantifer $\exists$ are the usual logical operators from
first-order logic. The operator $\dmd$ is a modal
operator (denoting ``possibility'').
Classical negation $\neg F$ is defined as $F \limp \bot$. Other
operators, i.e., $\top$, $\lor$, $\land$, $\forall$, and $\Box$,
can be defined via classical negation,
e.g., $\Box A = \dmd (A \limp \bot) \limp \bot$.
The connectives $\mand$ and
$\mimp$ correspond to separation conjunction and the ``magic wand''
from separation logic~\cite{reynolds2002}, and $\top^*$ is the multiplicative truth.

\begin{table*}[ht!]
\centering
\begin{tabular}{l@{\extracolsep{1cm}}l}
  \begin{tabular}{lcl}    
  $\Mcal, v,h \Vdash p(t_1,\cdots,t_n)$ & iff & $(t_1^\Mcal,\cdots,t_n^\Mcal)\in p^I$\\
$\Mcal, v,h \Vdash A\limp B$ & iff & $\Mcal, v,h\not \Vdash A$ or $\Mcal, v,h\Vdash B$\\
\end{tabular}
&
\begin{tabular}{lcl}
$\Mcal, v,h \Vdash \top^*$ & iff & $h = \epsilon$\\
$\Mcal, v,h \Vdash \bot$ & iff & never 
\end{tabular}\\
\multicolumn{2}{l}{
\hspace{-6px}
\begin{tabular}{lcl}
$\;\;\Mcal, v,h \Vdash A\mand B$ & iff & 
$h_1\circ h_2= h$ and $\Mcal,v, h_1\Vdash A$ and $\Mcal, v,h_2 \Vdash B$ for some $h_1$ and $h_2$\\
$\;\;\Mcal, v,h \Vdash A \mimp B$ & iff & 
for all $h_1$ and $h_2$, if $h\circ h_1=h_2$ and $\Mcal, v,h_1 \Vdash A$, then $\Mcal, v,h_2 \Vdash B.$
\end{tabular}
}\\
\begin{tabular}{lcl}
  $\Mcal, v,h \Vdash \exists x.A(x)$ & iff & $\exists d\in D. \Mcal, v[d/x],h\Vdash A(x)$\\
\end{tabular}
&
\\
\multicolumn{2}{l}{
\begin{tabular}{lcl}
  $\Mcal,v,h\Vdash \dmd A$ & iff & $\exists h_1\in H. \Mcal, v, h_1\Vdash A$\\  
  $\Mcal,v,h\Vdash t_1 = t_2$ & iff & $t_1^{\Mcal}$ and $t_2^{\Mcal}$ are the same element in $D.$\\
  $\Mcal,v,h\Vdash t_1 \apointsto t_2,\dots,t_k$ & iff & $\fpt_k(t_1^\Mcal,\dots,t_k^\Mcal) = h.$
\end{tabular}
}\\
\end{tabular}
\vspace{3px}
\caption{The semantics of $\foasl$.}
\label{tab:foasl_semantics}
\vspace{-0.5cm}
\end{table*}

The semantics of $\foasl$ is defined based on a \emph{separation
  algebra}, i.e., a commutative monoid $(H,\circ,\epsilon)$ where $H$
is a non-empty set, $\circ$ is a partial binary function $H\times H
\pfun H$ written infix, and $\epsilon\in H$ is the unit. This
separation algebra satisfies the following conditions, where `$=$' is
interpreted as `both sides undefined, or both sides defined and
equal':
\begin{description}[noitemsep]
\item[identity:] $\forall h\in H.\, h\circ\epsilon = h$.
\item[commutativity:] $\forall h_1,h_2\in H.\, h_1\circ h_2 = h_2\circ
  h_1$.
\item[associativity:] $\forall h_1,h_2,h_3\in H.\, h_1\circ(h_2\circ
  h_3) = (h_1\circ h_2)\circ h_3$.
\item[cancellativity:] $\forall h_1,h_2,h_3,h_4\in H.$ if $h_1\circ
  h_2 = h_3$ and $h_1\circ h_4 = h_3$ then $h_2 = h_4$.
\item[indivisible unit:] $\mbox{ if } h_1\circ h_2 = \epsilon \mbox{
  then } h_1 = \epsilon.$
\item[disjointness:] $\forall h_1,h_2\in H. \mbox{ if } h_1\circ h_1
  = h_2 \mbox{ then } h_1 = \epsilon.$
\item[cross-split:] if $h_1\circ h_2 = h_0$ and $h_3\circ h_4 = h_0$,
  then $\exists h_{13},h_{14},h_{23},h_{24}\in H$ such that $h_{13}\circ
  h_{14} = h_1$, $h_{23}\circ h_{24} = h_2$, $h_{13}\circ h_{23} =
  h_3$, and $h_{14}\circ h_{24} = h_4$.
\end{description}
Note that \textit{partial-determinism} is assumed since
$\circ$ is a partial function: for any $h_1,h_2,h_3,h_4\in H$, if
$h_1\circ h_2 = h_3$ and $h_1\circ h_2 = h_4$ then $h_3 = h_4$.

A $\foasl$ model is a tuple $\Mcal = (D,I,v,\Fcal,H,\circ,\epsilon)$
where $D$ is a non-empty domain, 
$I$ is an interpretation function mapping constant symbols to elements of $D$,
and predicate symbols, other than $=$ and $\apointsto$, to relations.
The function $v$ is a
valuation function 
mapping variables to $D$.  We use a set $\Fcal$ of functions to interpret
the abstract points-to predicates.
To each abstract points-to
predicate of arity $n$, we associate an $n$-argument  
total function $\fpt_n : D \times \cdots \times D \mapsto H \in \Fcal$.  
The tuple $(H,\circ,\epsilon)$ is a separation algebra.  
For a predicate symbol $p$ and a constant symbol $c$, 
we write $p^I$ and $c^I$, respectively, for their interpretations under $I$.
We write $t^\Mcal$ for the interpretation of term $t$ in the
model $\Mcal$. For a variable $x$, $x^\Mcal = v(x)$. A term $t$ is
\emph{closed} if it has no variables, and we write $t^I$ for the
interpretation of $t$ since it is independent of the valuation
function $v$.

A separation algebra $(H, \circ, \epsilon)$
can be seen as a Kripke frame, where $H$ is the set of worlds
and 
the (ternary) accessibility relation $R$ is defined as:
$R(h_1,h_2,h_3)$ iff $h_1 \circ h_2 = h_3.$ 
Modal operators are thus a natural extension to $\foasl$.

The semantics of $\foasl$ formulae are defined via Kripke style
notations in Table~\ref{tab:foasl_semantics}, where $\Mcal =
(D,I,v,\Fcal,H,\circ,\epsilon)$ is a $\foasl$ model, and $h,h_1,h_2
\in H$.  In the table, we write $v[c/x]$ to denote the valuation
function that may differ from $v$ only in the mapping of $x$, i.e.,
$v[c/x](x) = c$ and $v[c/x](y) = v(y)$ if $y \not = x.$ 
A $\foasl$ formula $A$ is true at $h$ in the model $\Mcal =
(D,I,v,\Fcal,$ $H,\circ,\epsilon)$ if $\Mcal,v,h\Vdash A.$
It is true in $\Mcal$ if it is true at some $h$ in $H$. 
A formula is valid if it is true in all models; a
formula is satisfiable if it is true in some model. A formula is
called a \emph{sentence} if it has no free variables.

Besides the first-order language, $\foasl$ has two main extensions
over $\psl$: the abstract points-to predicate $\apointsto$ and the
modality $\dmd$. We explain their intuitions here. 
In the concrete SL models, the points-to predicate $[a \mapsto b]$
is true only in a singleton heap that maps $a$ to $b.$
In our abstract semantics for $[a \apointsto b]$, we drop the
requirement that the heap must be singleton. Instead, we generalize this
by parameterizing the semantics with a function (the function $\fpt$ discussed earlier)
that associates values to (possibily non-singleton) heaps.
The predicate $[a \apointsto b]$ is true in a world $h$ iff 
$h$ is the image of $\fpt_2(a,b).$
As a consequence of this interpretation of $\apointsto$, we have
the following properties where $\vec t$ is a list of fields:
\begin{itemize}
\item {\em (Injectivity)} If $s \apointsto \vec t$ holds in both
  $h_1$ and $h_2$, then $h_1 = h_2$.
\item {\em (Totality)} For any $s$ and $\vec t$, there is some
  $h$ such that $s\apointsto \vec t$ holds in $h$.
\end{itemize}
The latter in particular is a consequence of the fact that functions
in $\Fcal$ are total functions.
We do not impose any other properties on $\apointsto$. For example,
we do not assume an invalid address $\nil$ such that $\nil \mapsto
\vec t$ must be false. The reason we cannot disprove $\nil \apointsto
\vec t$ is partly because 
we do not insist on a fixed interpretation of $\nil$ in our logic.
This \emph{does not} mean that $\nil \apointsto \vec t$ is
valid in our logic; it is only satisfiable.  We can strengthen
$\apointsto$ by adding more theories to it, including a formula to
force $\nil \apointsto \vec t$ to be unsatisfiable. See
Section~\ref{sec:axioms} and~\ref{sec:imp} for details.


To motivate the need for the modal operators, consider
an example to approximate,
in our framework, a separation logic 
where the points-to relation maps an address to a
multiset of addresses. In the binary case, one could formalize this
as: 
\begin{center}
  $F = \forall x,y,z. (x \mapsto y, z) \limp (x \mapsto z, y)$
\end{center}
We can encode this property as a rule in a labelled sequent calculus
as shown below left.  When generalising this to points-to of
$(n+1)$-arities, we will have to consider adding many variants of
rules that permute the right hand side of $\mapsto$, which is
what we are trying to avoid.  Alternatively, we can add the formula $F$ to the
antecedent of the sequent, and attach a label $l$ to $F$, and do a
forward-chaining on $l:F$, as shown below right, where $\Gamma,\Delta$ are sets of labelled formulae:
\begin{center}
\AxiomC{$\Gamma; l: (a \mapsto c,b) \vdash \Delta$}
\UnaryInfC{$\Gamma; l: (a \mapsto b,c) \vdash \Delta$}
\DisplayProof
\qquad
\AxiomC{$\cdots \qquad \Gamma, l:F, l: (a \mapsto c,b) \vdash \Delta$}
\UnaryInfC{$\Gamma, l:F, l:(a\mapsto b,c) \vdash \Delta$}
\DisplayProof
\end{center}
where the $\cdots$ are the instantiation of $x, y, z$ with $a,b,c$
respectively, and the discharge of the assumption $(x \mapsto y,z)$ of
$F$.  However, if $(a \mapsto b, c)$ in the conclusion is attached to
another label (world) $m$, we then have to add $m:F$ to the
sequent. In effect, we would have to add an infinite set of labelled
formulae of the form $k:F$ to the sequent to achieve the same effect
of the inference rule.  With modalities, we can simply use $l:\Box F$,
which would then allow $F$ to be used at any world in the antecedent
of the sequent.


\begin{example}
  \label{ex:reynolds}
  Consider Reynolds's semantics for separation
  logic~\cite{reynolds2002}, with an abstract points-to predicate of
  arity two.  This can be shown to be an instance of our abstract
  semantics, where the domain $D$ is the set of integers, $H$ is the
  set of heaps (i.e., finite partial maps from integers to integers),
  $\epsilon$ denotes the empty heap, and the function $\fpt_2$ is
  defined as
  $
  \fpt_2(a,b) = [a \mapsto b]
  $ where $[a \mapsto b]$ is the singleton heap, mapping $a$ to $b$.
  The operation $\circ$ on $H$ is defined as heap composition. It can
  be shown that $(H, \circ, \epsilon)$ forms a separation algebra.
  Note that if we relax the interpretation of $H$ to allow infinite
  heaps, $(H, \circ, \epsilon)$ is still a separation algebra, 
  which shows that our semantics may admit
  non-standard interpretations of separation logic.
\end{example}

\section{$\lsfoasl$: A Labelled Calculus for $\foasl$}
\label{sec:lsfoasl}

Let $\lvar$ be an infinite set of \emph{label variables}, the set
$\Lcal$ of \emph{labels} is $\lvar\cup\{\epsilon\}$, where
$\epsilon\not\in \lvar$ is a label constant.  We overload the notation
and write $h$ with subscripts as labels.  A function $\rho: \Lcal
\rightarrow H$ from labels to worlds is a {\em label mapping} iff it
satisfies $\rho(\epsilon) = \epsilon$, mapping the label constant
$\epsilon$ to the identity world of $H$.  A {\em labelled formula} is
a pair consisting of a label and a formula.  We write a labelled
formula as $h : A$, when $h$ is the label and $A$ is the formula of
the labelled formula. A {\em relational atom} is an expression of the
form $(h_1, h_2 \simp h_3)$, where $h_1,h_2$ and $h_3$ are labels,
this corresponds to $h_1\circ h_2 = h_3$ in the semantics.  A
relational atom is not a formula; rather it can be thought of as a
structural component of a sequent.  A \emph{sequent} takes the form
$\myseq{\Gcal}{\Gamma}{\Delta}$ where $\Gcal$ is a set of 
relational atoms, $\Gamma,\Delta$ are sets of labelled formulae, and
$;$ denotes set union. Thus $\Gamma;h:A$ is the union of $\Gamma$ and
$\{h:A\}$. The left hand side of a sequent is the \emph{antecedent}
and the right hand side is the \emph{succedent}.

We call our labelled proof system $\lsfoasl$. The logical rules of
$\lsfoasl$ are shown in Figure~\ref{fig:LS_FOASL}, structural rules
are in Figure~\ref{fig:LS_FOASL_str}. 
To simplify some rules, we introduce the notation $h_1 \sim h_2$ as an abbreviation of
$(\epsilon, h_1 \simp h_2).$ We use the notation $[t/x]$ to denote a
variable substitution, and similarly $[h'/h]$ for a label
substitution, where $h$ is a label variable. The equality rules, for
terms ($=_1$ and $=_2$) and labels ($\sim_1$ and $\sim_2$), are the
usual equality rules (see e.g., \cite{Troelstra96}). These rules allow
one to replace a term (label) with its equal anywhere in the
sequent. Note that in those rules, the replacement of terms (labels)
need not be done for all occurrences of equal terms; one can
replace just one occurrence or more. 
For example, below left is a valid instance of $=_2$.
This is because both the premise and the conclusion of the rules
are instances of the sequent below right:
\begin{center}
\AxiomC{$h:s = t; h_1 : p(t, s) \vdash h_2 : q(s,s)$}
\RightLabel{\tiny $=_2$}
\UnaryInfC{$h:s = t; h_1 : p(s,s) \vdash h_2 : q(s,s)$}
\DisplayProof
\qquad
$h: s = t; h_1:p(x,s) \vdash h_2 : q(s,s)$
\end{center}
i.e., the premise sequent is obtained from the above sequent with
substitution $[t/x]$, and the conclusion sequent with $[s/x].$ A
similar remark applies for label replacements in sequents affected via
$\sim_2.$ The rules $\apointsto_1$ and $\apointsto_2$ respectively capture
the injectivity and the totality properties 
of the underlying semantic function interpreting $\apointsto$.

\begin{figure*}[t]
\scriptsize
  \centering
\begin{tabular}{cc}
\multicolumn{2}{c}{
\AxiomC{}
\RightLabel{\tiny $id$}
\UnaryInfC{$\Gcal;\Gamma;h:A \vdash h:A;\Delta$}
\DisplayProof
  $\quad$
\AxiomC{$$}
\RightLabel{\tiny $\bot L$}
\UnaryInfC{$\Gcal;\Gamma; h:\bot \vdash \Delta$}
\DisplayProof
$\quad$
\AxiomC{$\Gcal;h \sim \epsilon;\Gamma \vdash \Delta$}
\RightLabel{\tiny $\top^* L$}
\UnaryInfC{$\Gcal;\Gamma;h:\top^* \vdash \Delta$}
\DisplayProof
$\quad$
\AxiomC{}
\RightLabel{\tiny $\top^* R$}
\UnaryInfC{$\Gcal;\Gamma \vdash \epsilon:\top^*;\Delta$}
\DisplayProof
}\\[15px]
\multicolumn{2}{c}{
\AxiomC{$\Gcal;\Gamma;h:A \vdash h:B;\Delta$}
\RightLabel{\tiny $\limp R$}
\UnaryInfC{$\Gcal;\Gamma \vdash h:A\limp B; \Delta$}
\DisplayProof
$\qquad$
\AxiomC{$\Gcal;\Gamma \vdash h:A;\Delta$}
\AxiomC{$\Gcal;\Gamma;h:B \vdash \Delta$}
\RightLabel{\tiny $\limp L$}
\BinaryInfC{$\Gcal;\Gamma;h:A\limp B \vdash \Delta$}
\DisplayProof
}\\[15px]
\multicolumn{2}{c}{
\AxiomC{$(h_1,h_2 \simp h_0);\Gcal;\Gamma;h_1:A;h_2:B \vdash \Delta$}
\RightLabel{\tiny $\mand L$}
\UnaryInfC{$\Gcal;\Gamma;h_0:A\mand B \vdash \Delta$}
\DisplayProof
$\qquad$
\AxiomC{$(h_1,h_0 \simp h_2);\Gcal;\Gamma;h_1:A \vdash h_2:B;\Delta$}
\RightLabel{\tiny $\mimp R$}
\UnaryInfC{$\Gcal;\Gamma \vdash h_0:A\mimp B;\Delta$}
\DisplayProof
}\\[15px]
\multicolumn{2}{c}{
\AxiomC{$(h_1,h_2 \simp h_0);\Gcal;\Gamma \vdash h_1:A;h_0:A\mand B;\Delta$}
\AxiomC{$(h_1,h_2 \simp h_0);\Gcal;\Gamma \vdash h_2:B;h_0:A\mand B;\Delta$}
\RightLabel{\tiny $\mand R$}
\BinaryInfC{$(h_1,h_2 \simp h_0);\Gcal;\Gamma \vdash h_0:A\mand B;\Delta$}
\DisplayProof
}\\[15px]
\multicolumn{2}{c}{
\AxiomC{$(h_1,h_0 \simp h_2);\Gcal;\Gamma;h_0:A\mimp B \vdash h_1:A;\Delta$}
\AxiomC{$(h_1,h_0 \simp h_2);\Gcal;\Gamma;h_0:A\mimp B; h_2:B \vdash \Delta$}
\RightLabel{\tiny $\mimp L$}
\BinaryInfC{$(h_1,h_0 \simp h_2);\Gcal;\Gamma;h_0:A\mimp B \vdash \Delta$}
\DisplayProof
}
\\[15px]
\multicolumn{2}{c}{
\AxiomC{$\Gcal;\Gamma;h:A(y)\vdash\Delta$}
\RightLabel{\tiny $\exists L$}
\UnaryInfC{$\Gcal;\Gamma;h:\exists x.A(x)\vdash\Delta$}
\DisplayProof
\qquad
\AxiomC{$\Gcal;\Gamma\vdash h:A(t);h:\exists x.A(x);\Delta$}
\RightLabel{\tiny $\exists R$}
\UnaryInfC{$\Gcal;\Gamma\vdash h:\exists x.A(x);\Delta$}
\DisplayProof
\qquad
\AxiomC{$\Gcal;\Gamma;h':A\vdash\Delta$}
\RightLabel{\tiny $\dmd L$}
\UnaryInfC{$\Gcal;\Gamma;h:\dmd A\vdash\Delta$}
\DisplayProof
}\\[15px]
\multicolumn{2}{c}{
\AxiomC{\myseq{\Gcal}{\Gamma;h:t=t}{\Delta}}
\RightLabel{\tiny $=_1$}
\UnaryInfC{$\myseq{\Gcal}{\Gamma}{\Delta}$}
\DisplayProof
$\quad$
\AxiomC{$\Gcal;h:s = t; \Gamma[t/x]\vdash \Delta[t/x]$}
\RightLabel{\tiny $=_2$}
\UnaryInfC{$\Gcal; h:s = t; \Gamma[s/x]\vdash\Delta[s/x]$}
\DisplayProof
\quad
\AxiomC{$\Gcal;\Gamma\vdash h':A;h:\dmd A;\Delta$}
\RightLabel{\tiny $\dmd R$}
\UnaryInfC{$\Gcal;\Gamma\vdash h:\dmd A;\Delta$}
\DisplayProof
}\\[15px]
\multicolumn{2}{c}{
  \AxiomC{$\myseq{\Gcal}{\Gamma;h:s \apointsto \vec t}{\Delta}$}
  \RightLabel{\tiny $\apointsto_1$}
  \UnaryInfC{$\myseq{\Gcal}{\Gamma}{\Delta}$}
  \DisplayProof
 $\quad$
  \AxiomC{$\myseq{\Gcal;h_1 \sim h_2}{\Gamma; h_1: s \apointsto \vec t; h_2 : s \apointsto \vec t}{\Delta}$}
  \RightLabel{\tiny $\apointsto_2$}
  \UnaryInfC{$\myseq{\Gcal}{\Gamma;h_1: s \apointsto \vec t; h_2: s \apointsto \vec t}{\Delta}$}
  \DisplayProof
}\\[15px]
\multicolumn{2}{l}{\textbf{Side conditions:}} \\
\multicolumn{2}{l}{
In $\mand L$ and $\mimp R$, the labels $h_1$ and $h_2$
do not occur in the conclusion.} \\
\multicolumn{2}{l}{In $\exists L$, $y$ is not free in the conclusion.
  \hspace{2cm}
In $\dmd L$, $h'$ does not occur in the conclusion.}\\
\multicolumn{2}{l}{In $\apointsto_1$, $h$ does not occur in the conclusion.}\\
\end{tabular}
\caption{Logical rules in $\lsfoasl$.}
\label{fig:LS_FOASL}
\end{figure*}

\begin{figure*}[t]
\scriptsize
\centering
\begin{tabular}{cc}
\multicolumn{2}{c}{
\AxiomC{$h \sim h;\Gcal; \Gamma \vdash \Delta$}
\RightLabel{\tiny $\sim_1$}
\UnaryInfC{$\Gcal; \Gamma \vdash \Delta$}
\DisplayProof
$\qquad$
\AxiomC{$h_1 \sim h_2;\Gcal[h_2/h];\Gamma[h_2/h] \vdash \! \Delta[h_2/h]$}
\RightLabel{\tiny $\sim_2$}
\UnaryInfC{$h_1\sim h_2;\Gcal[h_1/h]; \Gamma[h_1/h] \vdash \Delta[h_1/h]$}
\DisplayProof
}\\[15px]
\multicolumn{2}{c}{
\AxiomC{$(h_2,h_1 \simp h_0);(h_1,h_2 \simp h_0);\Gcal; \Gamma \vdash \Delta$}
\RightLabel{\tiny $E$}
\UnaryInfC{$(h_1,h_2 \simp h_0);\Gcal; \Gamma \vdash \Delta$}
\DisplayProof
$\qquad$
\AxiomC{$(h_1,h_1 \simp h_2); h_1 \sim \epsilon; \Gcal;\Gamma \vdash\Delta$}
\RightLabel{\tiny $D$}
\UnaryInfC{$(h_1,h_1\simp h_2);\Gcal;\Gamma\vdash\Delta$}
\DisplayProof
}\\[15px]
\multicolumn{2}{c}{
\AxiomC{$(h_3,h_5 \simp h_0);(h_2,h_4 \simp h_5);(h_1,h_2 \simp h_0);(h_3,h_4 \simp h_1);\Gcal;\Gamma \vdash \Delta$}
\RightLabel{\tiny $A$}
\UnaryInfC{$(h_1,h_2 \simp h_0);(h_3,h_4 \simp h_1);\Gcal;\Gamma \vdash \Delta$}
\DisplayProof
}\\[15px]
\multicolumn{2}{c}{
\AxiomC{$(h_1,h_2\simp h_0); h_0 \sim h_3; \Gcal;\Gamma \vdash \Delta$}
\RightLabel{\tiny $P$}
\UnaryInfC{$(h_1,h_2\simp h_0);(h_1,h_2\simp h_3);\Gcal;\Gamma\vdash\Delta$}
\DisplayProof
  $\qquad$
\AxiomC{$(h_1,\!h_2\!\simp h_0); h_2 \sim h_3; \Gcal;\Gamma \vdash\Delta$}
\RightLabel{\tiny $C$}
\UnaryInfC{$(h_1,\!h_2\!\simp h_0);(h_1,h_3\simp h_0);\Gcal;\Gamma\vdash\!\Delta$}
\DisplayProof
}\\[15px]
\multicolumn{2}{c}{
\AxiomC{$\myseq{(h_5,h_6\simp h_1);(h_7,h_8\simp h_2);(h_5,h_7\simp
  h_3);(h_6,h_8\simp h_4);(h_1,h_2\simp h_0);(h_3,h_4\simp h_0);\Gcal}{\Gamma}{\!\Delta}$}
\RightLabel{\tiny $CS$}
\UnaryInfC{$\myseq{(h_1,h_2\simp h_0);(h_3,h_4\simp h_0);\Gcal}{\Gamma}{\Delta}$}
\DisplayProof
}\\[15px]
\\
\multicolumn{2}{l}{\textbf{Side conditions:}}\\
\multicolumn{2}{l}{In $A$, the label $h_5$ does not occur in the conclusion.}\\
\multicolumn{2}{l}{In $CS$, the labels $h_5,h_6,h_7,h_8$ do not occur in the conclusion.}\\
\end{tabular}
\caption{Structural rules in $\lsfoasl$.}
\label{fig:LS_FOASL_str}
\end{figure*}


An \emph{extended model} $(\Mcal,\rho)$ is a $\foasl$
model $\Mcal$ equipped with a label mapping $\rho$.  A sequent
$\myseq{\Gcal}{\Gamma}{\Delta}$ is \emph{falsifiable} in an extended
model if: (1) every relational atom $(h_1,h_2\simp h_3)\in\Gcal$ is
true, i.e., $\rho(h_1) \circ \rho(h_2) = \rho(h_3)$; (2) every
labelled formula $h:A\in\Gamma$ is true, i.e., $\Mcal,v,\rho(h)\Vdash A$;
(3) every labelled formula $h':B\in\Delta$ is false, i.e.,
$\Mcal,v,\rho(h')\not\Vdash B$. A sequent is falsifiable if it is
falsifiable in some extended model.

To prove a formula $F$, we start from the sequent $\vdash h:F$ with an
arbitrary label $h\neq \epsilon$, and try to derive a closed
derivation by applying inference rules backwards from this sequent. A
derivation is closed if every branch can be closed by a rule with no
premises. The soundness of $\lsfoasl$ can be proved by arguing that
each rule preserves falsifiability upwards. The proof is given in
Appendix~\ref{app:sound}.


\begin{theorem}[Soundness]
  \label{thm:sound}
  For every $\foasl$ formula $F$, if $\vdash h:F$ is derivable in
  $\lsfoasl$ for any label $h$, then $F$ is a valid $\foasl$ formula.
\end{theorem}

\section{Counter-model Construction}
\label{sec:cmc}

We now give a counter-model construction for $\lsfoasl$ to show that
$\lsfoasl$ is complete w.r.t. $\foasl$. The proof here is motivated by
the completeness proof of the labelled sequent calculus and labelled
tableaux for $\psl$~\cite{hou2013b,wendling2012}, but this proof is
significantly more complex, as can be seen in the definition of
Hintikka sequent below, which has almost twice as many cases as the
previous work. The constructed model extends the non-classical logic
model in the previous work with a Herbrand model as in first-order
logic. For space reasons we only set up the stage here and delay the
full proofs to Appendix~\ref{app:comp}.

We define a notion of \emph{saturated sequent}, i.e., \emph{Hintikka
  sequent}, on which all possible rule instances in $\lsfoasl$ have
been applied.  In the following, we denote with $R$ a relational atom
or a labelled formula.

\begin{definition}[Hintikka sequent]
  \label{definition:hintikka_seq}
  Let $L$ be a $\foasl$ language and let $T$ be the set of closed
  terms in $L$. A labelled sequent $\myseq{\Gcal}{\Gamma}{\Delta}$,
  where $\Gamma,\Delta$ are sets of labelled sentences, is a {\em
    Hintikka sequent} w.r.t. $L$ if it satisfies the following
  conditions for any sentences $A,B$, any terms $t,t'$, and any labels
  $h,h_0,h_1,h_2,h_3,h_4,h_5,h_6,h_7$:
  \begin{enumerate}
    \small
\item If $h_1:A\in \Gamma$ and $h_2:A\in\Delta$ then
  $h_1 \sim h_2 \not \in \Gcal.$
\item $h : \bot \not \in \Gamma$.
\item If $h:\top^*\in \Gamma$ then
  $h \sim \epsilon\in\Gcal.$
\item If $h:\top^* \in\Delta$ then
  $h \sim \epsilon \not\in \Gcal.$
\item If $h:A\limp B\in \Gamma$ then $h:A\in \Delta$ or $h:B\in\Gamma.$
\item If $h:A\limp B\in \Delta$ then $h:A\in \Gamma$ and $h:B\in\Delta.$
\item If $h_0:A\mand B\in \Gamma$ then $\exists h_1,h_2\in\Lcal$ s.t. 
$(h_1,h_2\simp h_0)\in\Gcal$, $h_1:A\in\Gamma$ and $h_2:B\in\Gamma.$
\item If $h_3:A\mand B\in \Delta$ then $\forall h_0,h_1,h_2\in\Lcal$
  if $(h_1,h_2\simp h_0)\in\Gcal$ and $h_0 \sim h_3\in\Gcal$ then
  $h_1:A\in\Delta$ or $h_2:B\in\Delta.$
\item If $h_3:A\mimp B\in \Gamma$ then $\forall h_0,h_1,h_2\in\Lcal$ 
  if $(h_1,h_2\simp h_0)\in\Gcal$ and $h_2 \sim h_3\in\Gcal$, 
  then $h_1:A\in\Delta$ or $h_0:B\in\Gamma.$
\item If $h_2:A\mimp B\in \Delta$ then $\exists h_0,h_1\in\Lcal$ s.t. 
  $(h_1,h_2\simp h_0)\in\Gcal$, $h_1:A\in\Gamma$ and $h_0:B\in\Delta.$
\item If $h:\exists x.A(x) \in \Gamma$ then $h:A(t)\in\Gamma$ for some $t\in T$.
\item If $h:\exists x.A(x) \in \Delta$ then $h:A(t)\in\Delta$ for every $t\in T$.
\item If $h:\dmd A\in\Gamma$ then $\exists h_1\in\Lcal$ s.t. $h_1:A\in\Gamma$.
\item If $h:\dmd A\in\Delta$ then $\forall h_1\in\Lcal$, $h_1:A\in\Delta$.
\item For any $t\in T$, $\exists h\in\Lcal$ s.t. $h:t = t\in\Gamma$.
\item If $h_1:t = t' \in \Gamma$ and $h_2:A[t/x] \in \Gamma$ ($h_2 : A[t/x] \in \Delta$)
  then $h_2:A[t'/x]\in\Gamma$ (resp. $h_2 : A[t'/x] \in \Delta$).
\item For any label $h\in \Lcal$, $h \sim h \in\Gcal.$
\item If $h_1 \sim h_2 \in \Gcal$ and a relational atom or a labelled formula $R[h_1/h]\in\Gcal\cup\Gamma$
  (resp. $R[h_1/h]\in\Delta$),
  then $R[h_2/h]\in\Gcal\cup\Gamma$ (resp. $R[h_2/h]\in\Delta$).
\item If $(h_1,h_2\simp h_0)\in\Gcal$ then $(h_2,h_1\simp h_0)\in\Gcal.$
\item If $\{(h_1,h_2 \simp h_0);(h_3,h_4 \simp h_6);h_1 \sim h_6\}\subseteq\Gcal$ 
  then $\exists h_5\in\Lcal$. $\{(h_3,h_5 \simp h_0),(h_2,h_4 \simp h_5)\}\subseteq\Gcal$.
\item If $\{(h_1,h_2\simp h_0);(h_3,h_4\simp h_9);h_0 \sim h_9\}\subseteq\Gcal$
  then $\exists h_5,h_6,h_7,$ $h_8\in\Lcal$ s.t. $\{(h_5,h_6\simp h_1),(h_7,h_8\simp h_2),(h_5,h_7\simp h_3),(h_6,h_8\simp h_4)\}\subseteq\Gcal$.
\item For every abstract points-to predicate $\apointsto^k$ in the language
  and for any $t_1,\dots,t_k'\in T$, $\exists h\in\Lcal$ s.t. $h:t_1 \apointsto^k t_2,\dots,t_k \in \Gamma$.
  \item If $\{h_1:s \apointsto \vec t,h_2:s \apointsto \vec
    t\}\subseteq\Gamma$ then $h_1\sim h_2\in\Gcal$.
  \item If $\{(h_1,h_3\simp h_2),h_1\sim h_3\}\subseteq\Gcal$ then
    $h_1\sim \epsilon\in\Gcal$.
  \item If $\{(h_1,h_2\simp h_0),(h_4,h_5\simp h_3),h_1\sim
    h_4,h_2\sim h_5\}\subseteq\Gcal$ then $h_0\sim h_3\in\Gcal$.
  \item If $\{(h_1,\!h_2\!\simp h_0),(h_4,h_5\simp h_3),h_1\sim
    h_4,h_0\sim h_3\}\subseteq\Gcal$ then $h_2\sim h_5\in\Gcal$.
\end{enumerate}

\end{definition}

The next lemma shows that we can build an extended $\foasl$ model
$(\Mcal,\rho)$ where $\Mcal = (D,I,v,\Fcal,H,\circ,\epsilon)$ that
falsifies the Hintikka sequent $\myseq{\Gcal}{\Gamma}{\Delta}$.
The $D,I$ part is a Herbrand model as in first-order logic.
The construction of the monoid $(H,\circ,\epsilon)$ is similar to the one for
$\psl$~\cite{hou2013b}, where $H$ is the set of equivalent classes of
labels in the sequent. The interpretation of the predicate
$\apointsto$ is defined based the set of functions $\Fcal$. For each
$n$-ary predicate $\apointsto^n $, there is a function
$\fpt_n\in\Fcal$ defined as below:
  \begin{align*}
    \fpt_n (t_1,\cdots,t_n) = [h]_\Gcal \text{ iff } & h': t_1
    \apointsto^n t_2,\cdots, t_n \in \Gamma
     \text{ and } h\sim h'\in\Gcal.
  \end{align*}
  where $[h]_\Gcal$ is the class of labels equivalent to $h$ in
  $\Gcal$. $\Fcal$ is the set of all such functions. By Condition 22
  and 23 of the Hintikka sequent, each function in $\Fcal$ must be a
  total function. The full proof is in Appendix~\ref{app:comp}.

\begin{lemma}[Hintikka's Lemma]
  \label{lem:Hintikka_model}
  Suppose $L$ is a $\foasl$ language with a non-empty set of closed
  terms.  Every Hintikka sequent w.r.t. $L$ is falsifiable.
\end{lemma}

Then we show how to construct a Hintikka sequent for an unprovable
formula using the proof system $\lsfoasl$.
Unlike the usual
procedure, we have to consider the rules with no (or more than one)
principal formulae. To this end, we define a notion of \emph{extended
  formulae} as in the previous work~\cite{hou2013b}:
\begin{align*} 
\scriptsize
  \exfml{F} ::=
  & ~F~\mid~\equiv_1~\mid~\equiv_2~\mid~\mapsto_1~\mid~\mapsto_2~\mid~\Ebb~\mid~\Abb~\mid~\CSbb~\mid~
  \approx_1~\mid~\approx_2~\mid\\
  & ~\Pbb~\mid~\Cbb~\mid~\Dbb
\end{align*}
Here, $F$ is a $\foasl$ formula, the other symbols correspond to the
special rules in $\lsfoasl$. For example, $\equiv_1$ and $\equiv_2$
correspond to rules $=_1$ and $=_2$; $\mapsto_1$ and $\mapsto_2$
correspond to $\apointsto_1$ and $\apointsto_2$; $\approx_1$ and
$\approx_2$ correspond to $\sim_1$ and $\sim_2$. The saturation
procedure is performed according to a schedule, which is defined
below.
\begin{definition}[Schedule]
  A \emph{rule instance} is a tuple $(O,h,\exfml{F},R,$ $S,n)$, where
  $O$ is either $0$ (left) or $1$ (right), $h$ is a label, $\exfml{F}$
  is an extended formula, $R$ is a set of relational atoms such that
  $|R| \leq 2$, $S$ is a set of labelled formulae with $|S|\leq 2$,
  and $n$ is a natural number. Let $\Ical$ denote the set of all rule
  instances.  A {\em schedule} is a function from natural numbers
  $\mathbb{N}$ to $\Ical.$ A schedule $\phi$ is {\em fair} if for
  every rule instance $I$, the set $\{i \mid \phi(i) = I \}$ is
  infinite.
\end{definition}

It is easy to verify that a fair schedule must exist. This is proved
by checking that $\Ical$ is a countable set~\cite{wendling2012}, which
follows from the fact that $\Ical$ is a finite product of countable
sets. We fix a fair schedule $\phi$ for the following proofs. We
assume the set $\Lcal$ of labels is totally ordered and can be
enumerated as $h_0,h_1,h_2,\cdots$, where $h_0 = \epsilon$.
Similarly, we assume an infinite set of closed terms which can be
enumerated as $t_0,t_1,t_2,\cdots$, all of which are disjoint from the
terms in $F$. Suppose $F$ is an unprovable formula, we start from the
sequent $\vdash h_1:F$ and construct an underivable sequent as below.

\begin{definition}
\label{definition:sequent-series}
Let $F$ be a formula which is not provable in $\lsfoasl$.  We assume
that every variable in $F$ is bounded, otherwise we can rewrite $F$ so
that unbounded variables are universally quantified.
We construct a
series of finite sequents
$\langle \myseq{\Gcal_i}{\Gamma_i}{\Delta_i} \rangle_{i \in \Ncal}$
from $F$ where $\Gcal_1 = \Gamma_1 = \emptyset$ and $\Delta_1 =
a_1:F$.  Suppose $\myseq{\Gcal_i}{\Gamma_i}{\Delta_i}$ has been
defined, we define $\myseq{\Gcal_{i+1}}{\Gamma_{i+1}}{\Delta_{i+1}}$
in the sequel.  Suppose $\phi(i) = (O_i,h_i,\exfml{F}_i,R_i,S_i,n_i).$
When we use $n_i$ to select a term (resp. label) in a formula
(resp. relational atom), we assume the terms (resp. labels) are
ordered from left to right. If $n_i$ is greater than the number of
terms in the formula (labels in the relational atom), then no effect
is taken. We only show a few cases here, and display this rather
involved construction in Appendix~\ref{app:comp}.
\begin{itemize}
\item If $O_i = 0$, $\exfml{F}_i$ is a $\foasl$ formula $C_i = F_1\mand F_2$ and
  $h_i:C_i\in\Gamma_i$, then
$\Gcal_{i+1} =
  \Gcal_i\cup\{(h_{4i},h_{4i+1}\simp h_i)\}$, $\Gamma_{i+1} =
  \Gamma_i\cup\{h_{4i}:F_1,h_{4i+1}:F_2\}$, $\Delta_{i+1} = \Delta_i$.
\item If $\exfml{F}_i$ is $\equiv_1$ and $S_i = \{h_i:t_n = t_n\}$,
  where $n\leq i+1$, then $\Gcal_{i+1} = \Gcal_i$, $\Gamma_{i+1} =
  \Gamma_i\cup\{h_i:t_n = t_n\}$, and $\Delta_{i+1} = \Delta_i$.
\item If $\exfml{F}_i$ is $\equiv_2$ and $S_i = \{h:t = t',
  h':A[t/x]\}\subseteq\Gamma_i$, where $x$ is the $n_i$th term in $A$,
  then $\Gcal_{i+1} = \Gcal_i$, $\Gamma_{i+1} =
  \Gamma_i\cup\{h':A[t'/x]\}$, and $\Delta_{i+1} = \Delta_i$.
\item If $\exfml{F}_i$ is $\equiv_2$ and $S_i = \{h:t = t',
  h':A[t/x]\}$ where $h:t = t'\in\Gamma_i$, $h':A[t/x]\in\Delta_i$, and
  $x$ is the $n_i$th term in $A$. Then $\Gcal_{i+1} = \Gcal_i$,
  $\Gamma_{i+1} = \Gamma_i$, and $\Delta_{i+1} =
  \Delta_i\cup\{h':A[t'/x]\}$.
\end{itemize}
\end{definition}

The first rule shows how to use the $\mand L$ rule and how to deal
with fresh variables. The indexing of labels guarantees that the
choice of $h_{4i}$, $h_{4i+1}$, $h_{4i+2}$, $h_{4i+3}$ are always
fresh for the sequent $\myseq{\Gcal_i}{\Gamma_i}{\Delta_i}$.
Similarly, the term $t_{i+1}$ does not occur in the sequent
$\myseq{\Gcal_i}{\Gamma_i}{\Delta_i}$.  The second rule generates
an identity equality relation for the term $t_n$. The last two rules find a
formula $h':A$ in the antecedent and succedent respectively, and
replace $t$ with $t'$ in $A$. The construction in
Definition~\ref{definition:sequent-series} non-trivially extends a
similar construction of Hintikka CSS due to
Larchey-Wendling~\cite{wendling2012} and a similar one
in~\cite{hou2013b}.

We also borrow the notions of consistency and
finite-consistency from Larchey-Wendling's work~\cite{wendling2012}.
We say $\myseq{\Gcal'}{\Gamma'}{\Delta'} \subseteq
\myseq{\Gcal}{\Gamma}{\Delta}$ iff $\Gcal'\subseteq\Gcal$,
$\Gamma'\subseteq\Gamma$ and $\Delta'\subseteq\Delta$. A 
sequent $\myseq{\Gcal}{\Gamma}{\Delta}$ is \textit{finite} if
$\Gcal,\Gamma,\Delta$ are finite sets. Define
$\myseq{\Gcal'}{\Gamma'}{\Delta'} \subseteq_f
\myseq{\Gcal}{\Gamma}{\Delta}$ iff $\myseq{\Gcal'}{\Gamma'}{\Delta'}
\subseteq \myseq{\Gcal}{\Gamma}{\Delta}$ and
$\myseq{\Gcal'}{\Gamma'}{\Delta'}$ is finite. If
$\myseq{\Gcal}{\Gamma}{\Delta}$ is a finite sequent, it is
\textit{consistent} iff it does not have a derivation in $\lsfoasl$.
A (possibly infinite) sequent $\myseq{\Gcal}{\Gamma}{\Delta}$ is
\textit{finitely-consistent} iff every
$\myseq{\Gcal'}{\Gamma'}{\Delta'}\subseteq_f
\myseq{\Gcal}{\Gamma}{\Delta}$ is consistent.

We write $\Lcal_i$ for the set of labels occurring in the sequent
$\myseq{\Gcal_i}{\Gamma_i}{\Delta_i}$, and write $D_i$ for the set of
terms which are disjoint from those in $F$ in that sequent. Thus
$\Lcal_1 = \{a_1\}$ and $D_1 = \emptyset$. The following lemma states some properties of the
construction of the sequents $\myseq{\Gcal_i}{\Gamma_i}{\Delta_i}$.
\begin{lemma}
\label{lm:construction}
For any $i\in\mathcal{N}$, the following properties hold:
\begin{multicols}{2}
\begin{enumerate}
\item $\myseq{\Gcal_i}{\Gamma_i}{\Delta_i}$ has no derivation
\item $\Lcal_i\subseteq \{a_0, a_1,\cdots,a_{4i-1}\}$
\item $D_i\subseteq \{t_0,t_1,\cdots,t_i\}$
\item $\myseq{\Gcal_i}{\Gamma_i}{\Delta_i}\subseteq_f
  \myseq{\Gcal_{i+1}}{\Gamma_{i+1}}{\Delta_{i+1}}$
\end{enumerate}
\end{multicols}
\end{lemma}

Given the construction of the series of sequents in
Definition~\ref{definition:sequent-series}, we define a notion of a
{\em limit sequent} as the union of every sequent in the series.

\begin{definition}[Limit sequent]
\label{definition:lim_seq} 
Let $F$ be a formula unprovable in $\lsfoasl.$
The {\em limit sequent for $F$} is the sequent 
$\myseq{\Gcal^\omega}{\Gamma^\omega}{\Delta^\omega}$ 
where 
$\Gcal^\omega = \bigcup_{i\in\mathcal{N}}\Gcal_i$
and
$\Gamma^\omega = \bigcup_{i\in\mathcal{N}}\Gamma_i$
and
$\Delta^\omega = \bigcup_{i\in\mathcal{N}}\Delta_i$
and
where $\myseq{\Gcal_i}{\Gamma_i}{\Delta_i}$ is as defined in Def.\ref{definition:sequent-series}.
\end{definition}

The last step is to show that the limit sequent is a Hintikka sequent,
which gives rise to a counter-model of the formula that cannot be
proved.

\begin{lemma}
\label{lem:lim_hintikka}
If $F$ is a formula unprovable in $\lsfoasl$, then the limit sequent
for $F$ is a Hintikka sequent.
\end{lemma}

Now we can finish the completeness theorem: whenever a $\foasl$
formula has no derivation in $\lsfoasl$, there is an infinite
counter-model. The theorem states the contraposition.

\begin{theorem}[Completeness]
\label{thm:lsfoasl_complete}
If $F$ is valid in $\foasl$, then $F$ is provable in $\lsfoasl$.
\end{theorem}

\section{Theories for $\mapsto$ in Separation Logics}
\label{sec:axioms}

Our predicate $\apointsto$ admits more interpretations than
the standard $\mapsto$ predicate in SL heap model semantics. 
We can, however, approximate the behaviors of $\mapsto$ by
formulating additional properties of $\mapsto$ as logical theories.
We show next some of the theories for $\mapsto$ arising in
various SL semantics.


\subsection{Reynolds's semantics}
\label{sec:reynolds}

The $\mapsto$ predicate in Reynolds's semantics can be formalized
as follows, 
where the store $s$ is a total function from variables to values, 
and the heap $h$ is a finite partial function from addresses to values:
\begin{center}
$s,h\Vdash x \mapsto y$ iff $dom(h) = \{s(x)\}$ and $h(s(x)) = s(y)$.
\end{center}

Here we tackle the problem indirectly from the abstract separation
logic angle.
We give the following theories to approximate the
semantics of $\mapsto$ in SL:

\begin{enumerate}
\item $\Box \forall e_1,e_2. (e_1\mapsto e_2) \land \top^* \limp \bot$
  \hspace{1.5cm}
  \begin{minipage}{0.5\linewidth}
\item $\Box \forall e_1,e_2. (e_1\mapsto e_2) \limp \lnot (\lnot
  \top^* \mand \lnot \top^*)$
  \end{minipage}
\item $\Box \forall e_1,e_2,e_3,e_4. (e_1\mapsto e_2)\mand (e_3\mapsto
  e_4) \limp \lnot (e_1 = e_3)$
\item $\Box \forall e_1,e_2,e_3,e_4. (e_1\mapsto e_2)\land (e_3 \mapsto
  e_4) \limp (e_1 = e_3) \land (e_2 = e_4)$
\item $\Box \exists e_1 \forall e_2. \lnot ((e_1 \mapsto e_2) \mimp
  \bot)$
  \hspace{1.7cm}
  \begin{minipage}{0.5\linewidth}
    \item $\Box\forall e_1,e_2. (e_1\mapsto e_2) \limp (e_1\apointsto e_2)$
  \end{minipage}
\end{enumerate}
Note that the opposite direction $(e_1\apointsto e_2) \limp
(e_1\mapsto e_2)$ does not necessarily hold because $\apointsto$ is
weaker than $\mapsto$.
The above theories intend to capture the inference rules for $\mapsto$
in $\lssl$~\cite{hou2015}, the captured rules are given in
Figure~\ref{fig:LS_SL_pointer}. The first five formulae simulate the
rules $\mapsto L_1$, $\mapsto L_2$, $\mapsto L_3$, $\mapsto
L_4$, and $HE$ respectively. The rule $\mapsto L_5$ can be derived by
$\apointsto_2$ and Formula 6.

\begin{figure*}[t]
\scriptsize
\centering
\begin{tabular}{cc}
\multicolumn{2}{c}{
\AxiomC{}
\RightLabel{\tiny $\mapsto L_1$}
\UnaryInfC{$\Gcal;\Gamma;\epsilon:e_1\mapsto e_2\vdash\Delta$}
\DisplayProof
}\\[20px]
\multicolumn{2}{c}{
\AxiomC{$(\epsilon,h_0\simp h_0);\Gcal[\epsilon/h_1, h_0/h_2];\Gamma[\epsilon/h_1, h_0/h_2];h_0:e_1\mapsto e_2\vdash\Delta[\epsilon/h_1, h_0/h_2]$}
\alwaysNoLine
\UnaryInfC{$(h_0,\epsilon\simp h_0);\Gcal[\epsilon/h_2,h_0/h_1];\Gamma[\epsilon/h_2,h_0/h_1];h_0:e_1\mapsto e_2\vdash\Delta[\epsilon/h_2,h_0/h_1]$}
\alwaysSingleLine
\RightLabel{\tiny $\mapsto L_2$}
\UnaryInfC{$(h_1,h_2\simp h_0);\Gcal;\Gamma;h_0:e_1\mapsto e_2\vdash\Delta$}
\DisplayProof
}\\[20px]
\multicolumn{2}{c}{
\AxiomC{}
\RightLabel{\tiny $\mapsto L_3$}
\UnaryInfC{$(h_1,h_2\simp h_0);\Gcal;\Gamma;h_1:e\mapsto e_1;h_2:e\mapsto e_2\vdash\Delta$}
\DisplayProof
\qquad
\AxiomC{$\Gcal;\Gamma\theta;h:e_1\theta \mapsto e_2\theta \vdash\Delta\theta$}
\RightLabel{\tiny $\mapsto L_4$}
\UnaryInfC{$\Gcal;\Gamma;h:e_1\mapsto e_2;h:e_3\mapsto e_4\vdash\Delta$}
\DisplayProof
}\\[20px]
\multicolumn{2}{c}{
\AxiomC{$\myseq{\Gcal[h_1/h_2]}{\Gamma[h_1/h_2];h_1:e_1\mapsto e_2}{\Delta[h_1/h_2]}$}
\RightLabel{\tiny $\mapsto L_5$}
\UnaryInfC{$\myseq{\Gcal}{\Gamma;h_1:e_1\mapsto e_2;h_2:e_1\mapsto e_2}{\Delta}$}
\DisplayProof
\qquad
\AxiomC{$\myseq{(h_1,h_0\simp h_2);\Gcal}{\Gamma;h_1:e_1\mapsto e_2}{\Delta}$}
\RightLabel{\tiny $HE$}
\UnaryInfC{$\myseq{\Gcal}{\Gamma}{\Delta}$}
\DisplayProof
}\\[10px]
\multicolumn{2}{l}{\textbf{Side conditions:}} \\
\multicolumn{2}{l}{Each label being substituted cannot be $\epsilon$.
  \hspace{2cm}
In $\mapsto L_4$, $\theta = mgu(\{(e_1,e_3), (e_2,e_4)\}).$}\\
\multicolumn{2}{l}{In $HE$, $h_0$ occurs in conclusion, $h_1,h_2, e_1$ are fresh.}
\end{tabular}
\caption{Points-to rules in $\lssl$.}
\label{fig:LS_SL_pointer}
\end{figure*}

\begin{lemma}
The inference rules in Figure~\ref{fig:LS_SL_pointer} are admissible
in $\lsfoasl$ when Formula 1 $\sim$ 6 are assumed true.
\end{lemma}

The validity of Formula 1 to 6 w.r.t. Reynolds's SL model is easy to
check, the rationale is similar to the soundness of corresponding
rules in $\lssl$~\cite{hou2015}. Therefore Reynolds's SL is an
instance of our logic.

\begin{lemma}
Formula 1 $\sim$ 6 are valid in Reynolds's SL semantics.
\end{lemma}

The rules in Figure~\ref{fig:LS_SL_pointer} cover most of the rules
for $\mapsto$ in $\lssl$~\cite{hou2015}, but we have not found a way
to handle the following rule (with two premises):
\begin{center}
\AxiomC{$\myseq{(h_1,h_2\simp
    h_0);\Gcal}{\Gamma}{\Delta}$}
\noLine
\UnaryInfC{$\myseq{(h_3,h_4\simp h_1);(h_5,h_6\simp
    h_2);\Gcal}{\Gamma;h_3:e_1\mapsto e_2;h_5:e_1\mapsto
    e_3}{\Delta}$}
\RightLabel{\tiny $HC$}
\UnaryInfC{$\myseq{\Gcal}{\Gamma}{\Delta}$}
\DisplayProof
\end{center}
The rule $HC$ effectively picks two arbitrary heaps $h_1$ and $h_2$,
and does a case split of whether they can be combined or not. This
rule seems to require more expressiveness than our logic. However, the
above formulae cover most of properties about $\mapsto$ that existing
tools for SL can handle, including the treatments in~\cite{hou2015}
and those for symbolic heaps~\cite{berdine2005b}.

\subsection{Vafeiadis and Parkinson's SL}
\label{subsec:vp_sl}

Vafeiadis and Parkinson's
SL~\cite{vafeiadis2007} is almost the same as
Reynolds's definition, but they only consider values as
addresses. This is a common setting in many applications, such
as~\cite{Galmiche2005}. In this setting, the following formula is
valid:
$\top^* \limp \lnot ((e_1\mapsto e_2) \mimp \lnot (e_1 \mapsto e_2))$.
This formula, however, is invalid in Reynolds's SL. Obviously Formula
1 to 6 are valid in Vafeiadis and Parkinson's SL, thus their logic is
also an instance of our abstract logic. To cater for the
special feature, we propose a formula for ``total
addressability'':

\begin{enumerate}
\setcounter{enumi}{6}
\item $\forall e_1,e_2.\dmd (e_1 \mapsto e_2)$
\end{enumerate}

This formula ensures that there must exist a heap $(e_1 \mapsto e_2)$
no matter what values $e_1, e_2$ have.
This is sound because in Vafeiadis
and Parkinson's SL, $e_1$ must denote a valid address, thus $h$ with
$dom(h) = \{s(e_1)\}$ and $h(s(e_1)) = s(e_2)$, where $s$ is the
store, must be a legitimate function, which by definition is a
heap.

\subsection{Lee et al.'s SL}
\label{subsec:lee_sl}

Lee et al.'s proof system for SL corresponds to
a non-standard semantics (although they used Reynolds's semantics in
their paper)~\cite{Lee2013}. While there is not a reference of a formal
definition of their non-standard semantics, their inference rule
$\mimp Disj$ suggests that they forbid ``incompatible heaps''. For
example, if there exists a heap $e_1 \mapsto e_2$, then there shall
not exist another heap $(e_1\mapsto e_3)$, where $e_2 \neq e_3$.
Their $\mimp Disj$ rule can help derive the following formula, which
is invalid in Reynolds's SL:
\begin{align*}
  & (((e_1 \mapsto e_2) \mand \top ) \mimp \bot) \lor
  \label{fm:park_invalid}
  (((e_1 \mapsto e_3) \mand \top) \mimp \lnot((e_1 \mapsto e_2)\mimp \bot)) \lor (e_2 = e_3)
\end{align*}


If we assume that the above non-standard semantics conform with Reynolds's SL
in other aspects (as validated by Formula 1 to 6), then it
can be seen as a special instance of our abstract logic.
The compatibility property can then be formulated as follows:
\begin{enumerate}
\setcounter{enumi}{7}
\item $\forall e_1,e_2.
\dmd(e_1 \mapsto e_2) \limp \lnot(\exists e_3. \lnot(e_2 = e_3)
\land \dmd(e_1 \mapsto e_3))$
\end{enumerate}

With Formula 8 we can prove the invalid formula above.

\subsection{Thakur et al.'s SL}

There are SL variants that forbid heaps with cyclic lists, for
example, the one defined by Thakur et al.~\cite{thakur2014}.
Consequently,
the following two formulae are unsatisfiable in their SL:
\begin{center}
\begin{tabular}{c@{\hskip 2cm}c}
  $e_1\mapsto e_1$
  &
  $e_1\mapsto e_2 \mand e_2 \mapsto e_1$
\end{tabular}
\end{center}
To formulate this property, we first define a notion of a path: 
\begin{center}
$\forall e_1,e_2. \Box (path(e_1,e_2)  \equiv
e_1\mapsto e_2 \lor (\exists e_3. (e_1\mapsto e_3) \mand path(e_3,e_2)))$
\end{center}
where $\equiv$ denotes logical equivalence (bi-implication). 
Now the property of ``acyclism'' can be formulated as
\begin{enumerate}
  \setcounter{enumi}{8}
  \item $\forall e_1,e_2. \Box (path(e_1,e_2) \limp e_1\neq e_2)$
\end{enumerate}
which renders cyclic paths unsatisfiable in our logic, too.  Note that
since our proof system does not support inductive definitions,
we cannot force the
interpretation of $path$
to be
the least fixed point of its
definition.
We leave the incorporation of inductive definitions to future work.

\section{Implementation and Experiment}
\label{sec:imp}

Our theorem prover for $\foasl$ extends our previous prover for
Reynolds's 
SL~\cite{hou2015} with the ability to handle
(non-inductive) predicates and modalities. To speed up proof search, instead of
implementing $=_2$ and $\sim_2$, we use the following rules:
\begin{center}
  \AxiomC{$\myseq{\Gcal}{\Gamma[s/t]}{\Delta[s/t]}$}
  \RightLabel{\tiny $=_2'$}
  \UnaryInfC{$\myseq{\Gcal}{h:s=t;\Gamma}{\Delta}$}
  \DisplayProof
  \qquad
  \AxiomC{\myseq{\Gcal\theta}{\Gamma\theta}{\Delta\theta}}
  \RightLabel{\tiny $\sim_2'$}
  \UnaryInfC{$\myseq{h_1\sim h_2;\Gcal}{\Gamma}{\Delta}$}
  \DisplayProof\\[10px]
  where $\theta = [h_1/h_2]$ if $h_2\neq \epsilon$ and $\theta = [h_2/h_1]$ otherwise. 
\end{center}
These two rules can be shown to be interchangeable with $=_2$ and $\sim_2$.
One direction, i.e., showing that $=_2'$ and $\sim_2'$ can be derived in $\foasl$,
is straightforward. The other direction requires some further justification.
Let
$\lsfoasl'$ be $\lsfoasl$ with $=_2$ and $\sim_2$ replaced by $=_2'$
and $\sim_2'$ respectively, we then need to show that $=_2$ and
$\sim_2$ are admissible in $\lsfoasl'$.
To prove this, we follow a similar proof for free-equality rules for
first-order terms by Schroeder-Heister~\cite{Heister1994}. The key
part in that proof is in showing that provability is closed under substitutions.
In our setting, we
need to show that $\lsfoasl'$ is closed under both term substitutions
and label substitutions, which are stated below.

\begin{lemma}
If $\myseq{\Gcal}{\Gamma}{\Delta}$ is derivable in $\lsfoasl'$, then
so is $\myseq{\Gcal}{\Gamma[s/t]}{\Delta[s/t]}$ for any terms $s$ and
$t$.
\end{lemma}

\begin{lemma}
If $\myseq{\Gcal}{\Gamma}{\Delta}$ is derivable in $\lsfoasl'$, then
so is $\myseq{\Gcal[h_1/h_2]}{\Gamma[h_1/h_2]}{\Delta[h_1/h_2]}$ for
any label $h_1$ and label variable $h_2$.
\end{lemma}

Note that by restricting $h_2$ to a label variable, we forbid
$\epsilon$ to be substituted in the above lemma. These two lemmas
require induction on the height of derivations, and routine checks
confirm that they both hold. Then it is a corollary that $=_2$ and
$\sim_2$ are admissible in $\lsfoasl'$.

Since the heap model is widely used, our prover also includes useful
rules to reason in the heap model, such as the
derived rules in Figure~\ref{fig:LS_SL_pointer}.
But we currently have not included the $HC$ rule in our proof search
procedure. 
Since many applications of SL involve reasoning about
invalid addresses, such as $nil$, we also add a theory
to capture a simple aspect of the invalid address $nil$:
\begin{enumerate}
\setcounter{enumi}{9}
\item $\Box \forall \vec e.(nil \mapsto \vec e) \limp \bot$
\end{enumerate}
Since the current prover is an extension of our previous prover, it
builds in the inference rules for linked lists and binary trees for
reasoning about the \emph{symbolic heap} fragment of SL. It is also
capable of proving theorems used in verification of a tail-recursive
append function~\cite{maeda2011}, as shown in~\cite{hou2015}. However,
we do not exploit these aspects here.

\begin{table*}[t!]
  \centering
  \begin{tabular}{|l|@{\hskip 10px}l|l|}    
    \hline
    & Formula & Time\\
    \hline
    1 & $((\top \mimp (((k \mapsto c,d) \mimp (l \mapsto a,b)) \limp (l \mapsto a,b))) \limp (l \mapsto a,b))$ & $<$ 0.001s\\
    \hline
    2 & $((\exists x_2.((\exists x_1.((x_2 \mapsto x_1,b) \limp \bot)) \limp \bot)) \limp (\exists x_3.(x_3 \mapsto a,b)))$ & $<$ 0.001s\\
    \hline
    3 & $(((\top^* \limp \bot) \limp \bot) \limp ((\exists x_1.((x_1 \mapsto a,b) \mand \top)) \limp \bot))$ & $<$ 0.001s \\
    \hline
    4 & $((\exists x_3~x_2~x_1.(((x_3 \mapsto a,x_2) \mand (x_1 \mapsto c,d)) \land x_2 = x_1)) \limp$ &\\    
    & $(\exists x_5 ~x_4.((x_4 \mapsto c,d) \mand (x_5 \mapsto a,x_4))))$ & $<$ 0.001s\\
    \hline
    5 & $((((e_1 \mapsto e_2) \mand \top) \land (((e_3 \mapsto e_4) \mand \top) \land$ &\\
    & $(((e_5 \mapsto e_6) \mand \top) \land (\lnot (e_1 = e_3) \land (\lnot (e_1 =e_5) \land \lnot (e_3 =e_5)))))) \limp$ &\\
    & $(((e_1 \mapsto e_2) \mand ((e_3 \mapsto e_4) \mand (e_5 \mapsto e_6))) \mand \top))$ & 0.9s\\
    \hline
    6 & $((((e_1 \mapsto e_2) \mand \lnot((e_3 \mapsto e_4) \mand \top)) \land ((e_3 \mapsto e_4) \mand \top)) \limp e_1 = e_3)$ &  $<$ 0.001s\\
    \hline
    7 & $\lnot ((\lnot \top^* \mand  \lnot \top^*) \mimp \bot)$ & 0.0015s\\
    \hline
    8 & $((\lnot (((l_1 \mapsto p) \mand (l_2 \mapsto q)) \mimp (\lnot (l_3 \mapsto r)))) \limp$ &\\
    & $(\lnot ((l_1 \mapsto p) \mimp (\lnot (\lnot ((l_2 \mapsto q) \mimp (\lnot (l_3 \mapsto r))))))))$ & $<$ 0.001s\\
    \hline
    9 & $((\lnot ((l_1 \mapsto p) \mimp (\lnot (\lnot ((l_2 \mapsto q) \mimp (\lnot (l3 \mapsto r))))))) \limp$ &\\
    & $(\lnot (((l_1 \mapsto p) \mand (l_2 \mapsto q)) \mimp (\lnot (l_3 \mapsto r)))))$ & $<$ 0.001s\\
    \hline
    10 & $((\lnot ((lx \mapsto ly) \mimp (\lnot ((l1 \mapsto p) \mand (l2 \mapsto q))))) \limp$ &\\
    & $(\lnot ((\lnot ((\lnot ((lx \mapsto ly) \mimp (\lnot (l1 \mapsto p)))) \mand ((l2 \mapsto q) \land$ &\\
    & $ (\lnot (\exists x_1.((lx \mapsto x_1) \mand \top)))))) \land (\lnot ((\lnot ((lx \mapsto ly) \mimp$ &\\
    & $ (\lnot (l2 \mapsto q)))) \mand ((l1 \mapsto p) \land (\lnot (\exists x_2.((lx \mapsto x_2) \mand \top)))))))))$ & $<$ 0.001s\\
    \hline
    11 & $((\forall x_2 ~x_1.\dmd(x_2 \mapsto x_1)) \limp (\top^* \limp \lnot((e_1 \mapsto e_2) \mimp \lnot(e_1 \mapsto e_2))))$ & $<$ 0.001s\\
    \hline
    12 & $((\forall x_3 ~x_2.(\dmd(x_3 \mapsto x_2) \limp \lnot(\exists x_1.(\lnot (x2 =x1) \land \dmd(x_3 \mapsto x_1))))) \limp$ &\\
    & $((((e_1 \mapsto e_2) \mand \top) \mimp \bot) \lor ((((e_1 \mapsto e_3) \mand \top) \mimp \lnot((e_1 \mapsto e_2) \mimp \bot))$ &\\
    & $\lor e_2 =e_3)))$ & 0.0025s\\
    \hline
  \end{tabular}
  \vspace{10px}
  \caption{Experiment on selected formulae.}
  \vspace{-20px}
  \label{tab:exp1}
\end{table*}

We illustrate a list of formulae provable by our prover in
Table~\ref{tab:exp1}. Formulae 1 to 4 are examples drawn from Galmiche
and M\'ery's work on resource graph tableaux for
SL~\cite{galmiche2010}. Formula 5 is a property about overlaid data
structures: if the current heap contains $(e_1 \mapsto e_2)$ and $(e_3
\mapsto e_4)$ and $(e_5 \mapsto e_6)$, and they are pairwise distinct,
then the current heap contains the combination of the three heaps.
Formula 6 says that if the current heap can be split into two parts,
one is $(e_1 \mapsto e_2)$ and the other part does not contain $(e_3
\mapsto e_4)$, and the current heap contains $(e_3 \mapsto e_4)$, then
we deduce that $(e_3 \mapsto e_4)$ and $(e_1 \mapsto e_2)$ must be the
same heap, therefore $e_1 = e_3$. Formula 7 says that any heap can be
combined with a composite heap.  We give a derivation of formula 7 in
Appendix~\ref{app:ex_dev}. Formulae 8 to 10 are properties of
``septraction'' in SL with
Rely-Guarantee~\cite{vafeiadis2007}. Finally, formulae 11 and 12 show
that our prover can easily support reasoning about Vafeiadis and
Parkinson's SL (cf. Section~\ref{subsec:vp_sl}) and Lee et al.'s SL
(cf. Section~\ref{subsec:lee_sl}) by simply adding the corresponding
theories as assumptions. This is a great advantage over our previous
work where new rules have to be implemented to extend the ability of
the prover. To our knowledge most existing provers for SL cannot prove
the formulae in Table~\ref{tab:exp1}.  Examples of larger formulae
used in program verification can be found in the experiment of our
previous prover~\cite{hou2015}, upon which this prover is built.

\section{Conclusion}
\label{sec:conc}

This paper presents a first-order abstract separation logic with
modalities. This logic is rich enough to express formulae in
real-world applications such as program verification. We give a sound
and complete labelled sequent calculus for this logic. The
completeness of the finite calculus implies that our logic is
recursively enumerable. To deal with $\mapsto$, we give a set of
formulae to approximate the semantics of memory model. Of course, we
cannot fully simulate $\mapsto$, but we can handle most properties
about $\mapsto$ compared with existing tools for SL. Moreover, we can
prove numerous formulae that many existing tools for SL cannot
handle. The techniques discussed in this paper are demonstrated in a
rather flexible theorem prover which supports automated reasoning in
different SL variants without any change to the implementation. With
this foundation, one can simply add formulae as ``assumption'',
and prove theorems that cannot be proved in the base logic.

\section*{Acknowledgments}\label{sec:Acknowledgments}
This research is supported by the National Research Foundation, Prime Minister's Office, Singapore under its National Cybersecurity R\&D Program (Award No. NRF2014NCR-NCR001-30) and administered by the National Cybersecurity R\&D Directorate.


\bibliographystyle{plainurl}
\bibliography{main}

\newpage
\appendix

\section{Soundness of $\lsfoasl$}
\label{app:sound}

\textbf{Theorem~\ref{thm:sound}.} For every $\foasl$ formula $F$, if $\vdash h:F$ is derivable in
  $\lsfoasl$ for any label $h$, then $F$ is a valid $\foasl$ formula.

\begin{proof}
The soundness of most rules in $\lsfoasl$ is analogous to the same
rules in $\lspasl$~\cite{hou2013b}. Here we only show the rules
$\exists L$, $\exists R$, $\dmd L$, $\dmd R$, $\apointsto_1$, and
$\apointsto_2$. We prove the soundness of the rules by showing that if
the conclusion is falsifiable, then at least one premise is
falsifiable.
\begin{description}
\item[$\exists L$:]

  Assume the conclusion is falsifiable in an extended model
  $(\Mcal,\rho)$, we have $\Mcal,v,\rho(h)\Vdash \exists x.A$. So
  there is an element in $D$ such that by mapping $x$ to that
  element, $A$ is true at $h$. However, our valuation function $v$
  may not mention this element at all. We pick $y$ as that element,
  and let $v$ map $x$ to $y$. Since $y$ does not occur in the
  conclusion, this does not affect any existing mappings in $v$. Thus
  the mapping for all other variables remain the same, and the truth
  value of the reminder of the sequent does not change. With the
  extended valuation, $A[y/x]$ is true at $\rho(h)$, and the premise
  is falsifiable.

\item[$\exists R$:] Assume the conclusion is falsifiable in an
  extended model $(\Mcal,\rho)$, then $h:\exists x.A$ is false. That
  is, $\Mcal,v,\rho(h)\not\Vdash \exists x.A$, which means
  $\Mcal,v,\rho(h)\Vdash \forall x. \lnot A$. So there are no
  elements in $D$ that could possibly make $A$ true at
  $\rho(h)$. Therefore no matter what $y$ is, $A[y/x]$ is false at
  $\rho(h)$. So $\Mcal,v,\rho(h)\not\Vdash A[y/x]$. Hence the premise
  is falsifiable.

\item[$\dmd L$:] If the conclusion is falsifiable in some extended
  model $(\Mcal,\rho)$, we have $\Mcal,v,\rho(h)\Vdash\dmd A$. Thus
  there exists a world in $H$ such that $A$ under the valuation of $v$
  is true at that world. Let $\rho$ be extended with a fresh label
  $h'$ that maps to that world. This will not change the existing
  label mappings. Then we have $\Mcal,v,\rho(h')\Vdash A$. Thus the
  premise is falsifiable.

\item[$\dmd R$:] If the conclusion is falsifiable in some extended
  model $(\Mcal,\rho)$, then $\Mcal,v,\rho(h)\not\Vdash\dmd A$. This
  is equivalent to $\forall w\in H. \Mcal,v,w\not\Vdash A$. So for any
  mapping $\rho$ and any label $h'$, $\Mcal,v,\rho(h')\not\Vdash
  A$. We have that $h':A$ is false in the premise, thus the premise is
  falsifiable.

\item[$\apointsto_1$:] Suppose the conclusion is falsifiable in some extended
  model $(\Mcal,\rho)$.
  Suppose $\fpt$ is the semantic function associated with $\apointsto$. 
  Since $\fpt$ is a total function, and the domain is non-empty, 
  for any terms $s, \vec t$, there exists $h'\in H$ such that
  $\fpt(s^\Mcal, \vec t^\Mcal) = h'$. Let $\rho'$ be a label mapping
  such that $\rho'(h) = h'$ and $\rho'(h_1) = \rho(h_1)$ for any $h_1
  \not = h$. Since $h$ does not occur in the conclusion sequent, we
  have that $(\Mcal, \rho')$ satisfies all labelled formulae and
  relational atoms in the antecedent of the premise, and falsifies
  every formula in the succedent of the premise. Hence the premise is
  also fasifiable.
  
\item[$\apointsto_2$:] If the conclusion is falsifiable in some
  extended model $(\Mcal,\rho)$, then both $\Mcal,v,\rho(h_1)\Vdash s
  \apointsto \vec t$ and $\Mcal,v,\rho(h_2)\Vdash s \apointsto \vec t$
  hold. By the semantics of $\apointsto$, we have $\rho(h_1) =
  \rho(h_2)$. Thus $\epsilon \circ \rho(h_1) = \rho(h_2)$ holds, and
  $(\Mcal, \rho)$ satisfies $h_1\sim h_2$.
  So the premise is also falsifiable.
\end{description}
\vspace{-20px}
\qed
\end{proof}

\section{Completeness of $\lsfoasl$}
\label{app:comp}

 \textbf{Lemma~\ref{lem:Hintikka_model}.}
 Suppose $L$ is a $\foasl$ language with a non-empty set of closed
  terms.  Every Hintikka sequent w.r.t. $L$ is falsifiable.

\begin{proof}
  Let $\myseq{\Gcal}{\Gamma}{\Delta}$ be a Hintikka sequent
  w.r.t. $L$. We construct an extended model $(\Mcal,\rho)$ where
  $\Mcal = (D,I,v,\Fcal,H,\circ,\epsilon)$.

  The first order component $(D,I)$ is a Herbrand model. That is, 
  \begin{itemize}
  \item $D$ is exactly the non-empty set of closed terms of $L$.
  \item For each constant symbol $c$ in $L$, $c^I = c$.
  \end{itemize}
  It is then easy to verify that for each closed term $t$, $t^I =
  t$. As in the first order logic convention, the reminder of this
  proof will be confined to sentences and closed terms, so the
  valuation function $v$ does not matter here.

  The interpretation of the equality symbol $=$ is as follows:
  \begin{align*}
    t_1 = t_2
    \begin{cases}
      \text{ is true if $h:t_1 = t_2\in\Gamma$ for some $h\in\Lcal$;}\\
      \text{ is false otherwise.}
    \end{cases}
  \end{align*}
  We show that $=$ is indeed an equivalence relation: 
  \begin{description}
  \item[Reflexivity:] For any closed term $t\in D$, by condition 15 of
    Definition~\ref{definition:hintikka_seq}, $\exists h\in\Lcal$
    such that $h:t = t\in\Gamma$. Thus $t = t$.
  \item[Symmetry:] Suppose $t_1 = t_2$ holds, then there is some
    $h\in\Lcal$ such that $h:t_1 = t_2\in\Gamma$. By reflexivity, $t_1
    = t_1$ holds, and $h':t_1 = t_1\in\Gamma$ for some
    $h'\in\Lcal$. Consider $h':t_1 = t_1$ as an instance of $h':x =
    t_1[t_1/x]$, Condition 16 in
    Definition~\ref{definition:hintikka_seq} guarantees that $h':t_2 =
    t_1\in\Gamma$. Thus $t_2 = t_1$.
  \item[Transitivity:] Suppose $t_1 = t_2$ and $t_2 = t_3$ both
    hold. There must be some $h,h'\in\Lcal$ such that $\{h:t_1
    =t_2,h':t_2 = t_3\}\subseteq\Gamma$.
    Since $h:t_1
    = t_2$ is an instance of $h:t_1 = x[t_2/x]$, by condition 16 in
    Definition~\ref{definition:hintikka_seq}, $h:t_1 = t_3\in\Gamma$,
    thus $t_1 = t_3$.
  \end{description}

  Similarly, the relation $\sim$ on labels is also an equivalence
  relation. Given a set of relational atoms $\Gcal$, the relation
  $\sim$ partitions the set $\Lcal$ of labels into equivalence classes
  $[h]_\Gcal$ for each label $a \in \Lcal$:
$$[h]_\Gcal = \{ h' \in \Lcal \mid h \sim h'\in\Gcal \}.$$

  The other components of the extended model are defined as:
  \begin{itemize}
  \item $H = \{[h]_\Gcal \mid h\in\Lcal\}$.
  \item $[h_1]_\Gcal \circ [h_2]_\Gcal = [h_0]_\Gcal$ iff $\exists
    h_0',h_1',h_2'$ s.t. $\{(h_1',h_2'\simp h_0');h_0\sim h_0';h_1\sim
    h_1';h_2\sim h_2'\}\subseteq\Gcal$.
  \item $\epsilon = [\epsilon]_\Gcal$.
  \item $\rho(h) = [h]_\Gcal$ for every $h\in\Lcal$.
  \end{itemize}

  The interpretation of the predicate $\apointsto$ is defined based
  the set of functions $\Fcal$. For each $n$-ary predicate
  $\apointsto^n $, there is a function $\fpt_n\in\Fcal$ defined as
  below:
  \begin{align*}
    \fpt_n (t_1,\cdots,t_n) = [h]_\Gcal \text{ iff } & h': t_1
    \apointsto^n t_2,\cdots t_n \in \Gamma\\
    & \text{ and } h\sim h'\in\Gcal.
  \end{align*}
  $\Fcal$ is the set of all such functions. By Condition 22 and 23, each
  function in $\Fcal$ must be a total function.
  
  Finally, the interpretation for the other relation symbols are given
  below:
  \begin{itemize}
  \item For each $n$-ary relation symbol $R$ in $L$ and $n$ closed
    terms $t_1,\cdots,t_n$ in $D$, $R^I(t_1,\cdots,t_n)$ holds in the
    world $[h]_\Gcal\in H$ iff $h':R(t_1,\cdots,t_n)\in \Gamma$ and $h
    \sim h'\in\Gcal$.
  \end{itemize}

  In the following we will drop the subscript $\Gcal$ in $[h]_\Gcal$
  for readability. It can be checked that conditions 17, 19, 20, 21 in
  Definition~\ref{definition:hintikka_seq} ensure identity,
  commutativity, associativity, and cross-split respectively in the
  semantics. Furthermore, disjointness, partial-determinism,
  cancellativity are covered by Condition 24, 25, 26
  respectively. Condition 24 also implies indivisible unit. A similar
  proof to these can be found in~\cite{hou2013b}. Here we give three
  cases that are not considered in the previous work.
  \begin{description}
  \item[Cross-split:] Assume $[h_1] \circ [h_2] = [h_0]$ and $[h_3]
    \circ [h_4] = [h_0]$ hold. By the above construction, there must
    be some $(h_1',h_2'\simp h_0')\in\Gcal$ and $(h_3',h_4'\simp
    h_0'')\in\Gcal$ such that $[h_1] = [h_1']$, $[h_2] = [h_2']$,
    $[h_3] = [h_3']$, $[h_3] = [h_4']$, and $[h_0] = [h_0'] = [h_0'']$
    hold. This means that $h_0' \sim h_0''\in\Gcal$. By condition 21 in
    Definition~\ref{definition:hintikka_seq}, there exist
    $h_5,h_6,h_7,h_8\in\Lcal$ such that $\{(h_5,h_6\simp
    h_1'),(h_7,h_8\simp h_2'),(h_5,h_7\simp h_3'),(h_6,h_8\simp
    h_4')\}\subseteq\Gcal$. Again, by construction of the model, we
    have $[h_5]\circ [h_6] = [h_1']$, $[h_7]\circ [h_8] = [h_2']$,
    $[h_5]\circ [h_7] = [h_3']$, $[h_6]\circ [h_8] = [h_4']$, from
    which we obtain the cross-split property in the semantics.

  \item[Disjointness:] For any two worlds $[h_1],[h_2]\in H$, if
    $[h_1]\circ [h_1] = [h_2]$, then there must be some
    $(h_1',h_1''\simp h_2')\in\Gcal$ such that $[h_1] = [h_1'] =
    [h_1'']$ and $[h_2] = [h_2']$, hence $h_1' \sim h_1''\in\Gcal$.
    By Condition 24, $h_1'\sim \epsilon\in\Gcal$. Since $\sim$ is an
    equivalence relation, we have $h_1 \sim \epsilon$.

  \item[Indivisible unit:] This is a direct consequence of
    disjointness~\cite{larcheywendling2014}.
  \end{description}

  Therefore the $(H,\circ,\epsilon)$ part of the extended model is
  indeed a separation algebra as defined in
  Section~\ref{sec:foasl}. The reminder of this proof shows that the
  Hintikka sequent $\myseq{\Gcal}{\Gamma}{\Delta}$ is falsifiable in
  the constructed model, this includes the following:
  \begin{itemize}
  \item If $(h_1,h_2,\simp h_0)\in\Gcal$ then $[h_1]\circ[h_2] = [h_0]$.
  \item If $h:A\in\Gamma$ then $\Mcal,v,[h]\Vdash A$.
  \item If $h:A\in\Delta$ then $\Mcal,v,[h]\not\Vdash A$.
  \end{itemize}
  The first item follows from the construction of the model. In the
  sequel we show the last two items simultaneously by induction on the
  size of sentence $A$.

  \noindent {\bf Base cases:}
    \begin{itemize}
    \item Suppose an atomic labelled formula $h:R(t_1,\cdots,t_n)$ is
      in $\Gamma$. We must show that $(t_1^\Mcal,\cdots,t_n^\Mcal)\in
      R^I$ in the world $[h]$. Since $R(t_1,\cdots,t_n)$ is a
      sentence, $t_1,\cdots,t_n$ are closed terms. Thus each
      $t_i^\Mcal$ is actually $t_i^I$, which equals $t_i$. By the
      definition of our model, $R^I(t_1,\cdots,t_n)$ holds in the
      world $[h]$. So $\Mcal,v,[h]\Vdash R(t_1,\cdots,t_n)$ regardless
      of the valuation function $v$.

    \item Suppose $h:R(t_1,\cdots,t_n)\in\Delta$, we show that
      $\Mcal,v,h\not\Vdash R(t_1,\cdots,t_n)$. Assume the opposite,
      i.e., $R^I(t_1,\cdots,t_n)$ is true in $[h]$, then there must be
      some $h':R(t_1,\cdots,t_n)\in\Gamma$ and $h \sim h'\in\Gcal$. But
      this contradicts condition 1 in
      Definition~\ref{definition:hintikka_seq}. Thus
      $\Mcal,v,h\not\Vdash R(t_1,\cdots,t_n)$.\\

      Note that the cases for $\apointsto$ are just special cases of
      the above by the definition of $\fpt\in\Fcal$.
    \end{itemize}
  {\bf Inductive cases:}
   We do a case analysis on the main
    connective/quantifier/modality of the sentence $A$. The cases for
    connectives in propositional abstract separation logic are proved
    in~\cite{hou2013b}. Here we show the cases for $\exists$, $\dmd$, which correspond to conditions 11, 12, 13, 14 respectively in Definition~\ref{definition:hintikka_seq}.
    \begin{itemize}
    \item If $h:\exists x.A(x)\in\Gamma$, then $h:A(t)\in\Gamma$ for
      some closed term $t$ in $L$. Since $D$ is the set of closed
      terms in $L$, there exists some $t\in D$ such that when the
      valuation function $v$ maps $x$ to $t$, we have
      $\Mcal,v[t/x],[h]\Vdash A(x)$ by the induction hypothesis. Thus
      $\Mcal,v,[h]\Vdash \exists x.A(x)$.
    \item If $h:\exists x.A(x)\in\Delta$, then $h:A(t)\in\Delta$ for
      every closed term $t\in L$. Since $D$ is the set of closed terms
      in $L$, for any $t\in D$, if the valuation function $v$ maps $x$
      to $t$, we have $\Mcal,v[t/x],[h]\not\Vdash A(x)$ by the
      induction hypothesis. Thus $\Mcal,v,[h]\not\Vdash \exists
      x.A(x)$.
    \item If $h:\dmd A\in\Gamma$, then $\exists h'\in\Lcal$ such that
      $h':A\in\Gamma$. By the induction hypothesis, there is some
      $[h']\in H$ such that $\Mcal,v,[h']\Vdash A$. Therefore
      $\Mcal,v,[h]\Vdash \dmd A$.
    \item If $h:\dmd A\in\Delta$, then $\forall h'\in\Lcal$,
      $h':A\in\Delta$. By the induction hypothesis,
      $\Mcal,v,[h']\not\Vdash A$ for any $[h']\in H$. So
      $\Mcal,v,[h]\not\Vdash \dmd A$.
    \end{itemize}

\vspace{-17px}
\qed
\end{proof}

\vspace{10px}
\noindent \textbf{Definition~\ref{definition:sequent-series}.}
Let $F$ be a formula which is not provable in $\lsfoasl$.  We assume
that every variable in $F$ is bounded, otherwise we can rewrite $F$ so
that unbounded variables are universally quantified. It can be checked
that this will not affect the validity of the formula. We construct a
series of finite sequents
$\langle \myseq{\Gcal_i}{\Gamma_i}{\Delta_i} \rangle_{i \in \Ncal}$
from $F$ where $\Gcal_1 = \Gamma_1 = \emptyset$ and $\Delta_1 =
a_1:F$.  Suppose $\myseq{\Gcal_i}{\Gamma_i}{\Delta_i}$ has been
defined, we define $\myseq{\Gcal_{i+1}}{\Gamma_{i+1}}{\Delta_{i+1}}$
in the sequel.  Suppose $\phi(i) = (O_i,h_i,\exfml{F}_i,R_i,S_i,n_i).$
When we use $n_i$ to select a term (resp. label) in a formula
(resp. relational atom), we assume the terms (resp. labels) are
ordered from left to right. If $n_i$ is greater than the number of
terms in the formula (labels in the relational atom), then no effect
is taken.

\begin{itemize}
\item If $O_i = 0$, $\exfml{F}_i$ is a $\foasl$ formula $C_i$ and
  $h_i:C_i\in\Gamma_i$:
\begin{itemize}
\item If $C_i = F_1\limp F_2$. If there is no derivation for
  $\myseq{\Gcal_i}{\Gamma_i}{h_i:F_1;\Delta_i}$ then $\Gamma_{i+1} =
  \Gamma_i$, $\Delta_{i+1} = \Delta_i\cup\{h_i:F_1\}$. Otherwise
  $\Gamma_{i+1} = \Gamma_i\cup\{h_i:F_2\}$, $\Delta_{i+1} =
  \Delta_i$. In both cases, $\Gcal_{i+1} = \Gcal_i$.
\item If $C_i = \top^*$, then $\Gcal_{i+1} =
  \Gcal_i\cup\{(\epsilon,h_i\simp \epsilon)\}$, $\Gamma_{i+1} =
  \Gamma_i$, $\Delta_{i+1} = \Delta_i$.
\item If $C_i = F_1\mand F_2$, then $\Gcal_{i+1} =
  \Gcal_i\cup\{(h_{4i},h_{4i+1}\simp h_i)\}$, $\Gamma_{i+1} =
  \Gamma_i\cup\{h_{4i}:F_1,h_{4i+1}:F_2\}$, $\Delta_{i+1} = \Delta_i$.
\item If $C_i = F_1\mimp F_2$ and $R_i = \{(h,h_i\simp
  h'')\}\subseteq\Gcal_i$.  If
  $\myseq{\Gcal_i}{\Gamma_i}{h:F_1;\Delta_i}$ has no derivation, then
  $\Gamma_{i+1} = \Gamma_i$, $\Delta_{i+1} =
  \Delta_i\cup\{h:F_1\}$. Otherwise $\Gamma_{i+1} =
  \Gamma_i\cup\{h'':F_2\}$, $\Delta_{i+1} = \Delta_i$. In both cases,
  $\Gcal_{i+1} = \Gcal_i$.
\item If $C_i = \exists x.F(x)$, then $\Gcal_{i+1} = \Gcal_i$,
  $\Gamma_{i+1} = \Gamma_i\cup\{h_i:F(t_{i+1})\}$, and $\Delta_{i+1} =
  \Delta_i$.
\item If $C_i = \dmd F$, then $\Gcal_{i+1} = \Gcal_i$, $\Gamma_{i+1} =
  \Gamma_i\cup\{h_{4i}:F\}$, and $\Delta_{i+1} = \Delta_i$.
\end{itemize}

\item If $O_i = 1$, $\exfml{F}_i$ is a $\foasl$ formula $C_i$, and
  $h_i:C_i\in\Delta$:
\begin{itemize}
\item If $C_i = F_1\limp F_2$, then $\Gamma_{i+1} =
  \Gamma\cup\{h_i:F_1\}$, $\Delta_{i+1} = \Delta_i\cup\{h_i:F_2\}$,
  and $\Gcal_{i+1} = \Gcal_i$.
\item $C_i = F_1\mand F_2$ and $R_i = \{(h,h'\simp
  h_i)\}\subseteq\Gcal_i$. If
  $\myseq{\Gcal_i}{\Gamma_i}{h:F_1;\Delta_i}$ has no derivation, then
  $\Delta_{i+1} = \Delta_i\cup\{h:F_1\}$. Otherwise $\Delta_{i+1} =
  \Delta_i\cup\{h':F_2\}$. In both cases, $\Gcal_{i+1} = \Gcal_i$ and
  $\Gamma_{i+1} = \Gamma_i$.
\item If $C_i = F_1\mimp F_2$, then $\Gcal_{i+1} =
  \Gcal_i\cup\{(h_{4i},h_i\simp h_{4i+1})\}$, $\Gamma_{i+1} =
  \Gamma_i\cup\{h_{4i}:F_1\}$, and $\Delta_{i+1} =
  \Delta_i\cup\{h_{4i+1}:F_2\}$.
\item If $C_i = \exists x.F(x)$ and $S_i =\{h_i:t_n = t_n\}$, where $n\leq
  i+1$, then $\Gcal_{i+1} = \Gcal_i$, $\Gamma_{i+1} = \Gamma_i$,
  $\Delta_{i+1} = \Delta_i\cup\{h_i:F(t_n)\}$.
\item If $C_i = \dmd F$ and $R_i = \{(h_n,\epsilon\simp h_n)\}$, where
  $n\leq 4i+3$, then $\Gcal_{i+1} = \Gcal_i$, $\Gamma_{i+1} =
  \Gamma_i$, and $\Delta_{i+1} = \Delta_i\cup\{h_n:F\}$.
\end{itemize}
\item If $\exfml{F}_i
  \in\{\equiv_1,\equiv_2,\mapsto_1,\mapsto_2,\Ebb,\Abb,\CSbb,\approx_1,\approx_2,\Pbb,\Cbb,\Dbb\}$,
  we proceed as follows:
\begin{itemize}    
\item If $\exfml{F}_i$ is $\equiv_1$ and $S_i = \{h_i:t_n = t_n\}$,
  where $n\leq i+1$, then $\Gcal_{i+1} = \Gcal_i$, $\Gamma_{i+1} =
  \Gamma_i\cup\{h_i:t_n = t_n\}$, and $\Delta_{i+1} = \Delta_i$.
\item If $\exfml{F}_i$ is $\equiv_2$ and $S_i = \{h:t = t',
  h':A[t/x]\}\subseteq\Gamma_i$, where $x$ is the $n_i$th term in $A$,
  then $\Gcal_{i+1} = \Gcal_i$, $\Gamma_{i+1} =
  \Gamma_i\cup\{h':A[t'/x]\}$, and $\Delta_{i+1} = \Delta_i$.
\item If $\exfml{F}_i$ is $\equiv_2$ and $S_i = \{h:t = t',
  h':A[t/x]\}$ where $h:t = t'\in\Gamma_i$, $h':A[t/x]\in\Delta_i$, and
  $x$ is the $n_i$th term in $A$. Then $\Gcal_{i+1} = \Gcal_i$,
  $\Gamma_{i+1} = \Gamma_i$, and $\Delta_{i+1} =
  \Delta_i\cup\{h':A[t'/x]\}$.
\item If $\exfml{F}_i$ is $\mapsto_1$ and $S_i = \{h_i:e_1\apointsto
  \vec e\}$ where $\{e_1,\vec e\}\subseteq \{t_1,\cdots, t_{i+1}\}$. Then
  $\Gcal_{i+1} = \Gcal_i$, $\Gamma_{i+1} =
  \Gamma_i\cup\{h_{4i}:e_1\apointsto \vec e\}$, and $\Delta_{i+1} =
  \Delta_i$.
\item If $\exfml{F}_i$ is $\mapsto_2$ and $S_i = \{h:s\apointsto \vec
  e,h':s\apointsto \vec e\}\subseteq\Gamma_i$, then $\Gcal_{i+1} =
  \Gcal_i\cup \{h\sim h'\}$, $\Gamma_{i+1} = \Gamma_i$ and
  $\Delta_{i+1} = \Delta_i$.
\item If $\exfml{F}_i = \Ebb$, $R_i = \{(h,h'\simp h'')\}\subseteq\Gcal_i$, then
  $\Gcal_{i+1}=\Gcal_i\cup\{(h',h\simp h'')\}$, $\Gamma_{i+1} =
  \Gamma_i$, $\Delta_{i+1} = \Delta_i$.
\item If $\exfml{F}_i = \Abb$, $R_i = \{(h,h'\simp h''),(w,w'\simp
  h)\}\subseteq\Gcal_i$, then
  $\Gcal_{i+1}=\Gcal_i\cup\{(w,h_{4i}\simp h''),(h',w'\simp h_{4i})\}$,
  $\Gamma_{i+1} = \Gamma_i$, $\Delta_{i+1} = \Delta_i$.
\item If $\exfml{F}_i = \CSbb$, $R_i = \{(h,h'\simp h''),(w,w'\simp
  h'')\}\subseteq\Gcal_i$, then
  $\Gcal_{i+1} = \Gcal_i\cup\{(h_{4i},h_{4i+1}\simp
  h),(h_{4i},h_{4i+2}\simp w),(h_{4i+2},h_{4i+3}\simp
  h'),(h_{4i+1},h_{4i+3}\simp w')\}$, $\Gamma_{i+1} = \Gamma_i$,
  $\Delta_{i+1} = \Delta_i$.
\item If $\exfml{F}_i$ is $\approx_1$, $R_i = \{(\epsilon,h_n\simp
  h_n)\}$, where $n \leq 4i+3$, then
  $\Gcal_{i+1}=\Gcal_i\cup\{h_n\sim h_n\}$, $\Gamma_{i+1}
  = \Gamma_i$, $\Delta_{i+1} = \Delta_i$.
\item If $\exfml{F}_i$ is $\approx_2$ and $R_i = \{h' \sim
  h'',R[h'/h]\}\subseteq\Gcal_i$, where $h$ is the $n_i$th label in
  $R$. Then $\Gcal_{i+1} = \Gcal_i\cup\{R[h''/h]\}$, $\Gamma_{i+1} =
  \Gamma_i$, and $\Delta_{i+1} = \Delta_i$.
\item If $\exfml{F}_i$ is $\approx_2$, $R_i = \{h \sim
  h'\}\subseteq\Gcal_i$, and $S_i =
  \{w:A[h/w]\}\subseteq\Gamma_i$. Then $\Gcal_{i+1} = \Gcal_i$,
  $\Gamma_{i+1} = \Gamma_i\cup \{w:A[h'/w]\}$, and $\Delta_{i+1} =
  \Delta_i$.
\item If $\exfml{F}_i$ is $\approx_2$, $R_i = \{h \sim
  h'\}\subseteq\Gcal_i$, and $S_i =
  \{w:A[h/w]\}\subseteq\Delta_i$. Then $\Gcal_{i+1} = \Gcal_i$,
  $\Gamma_{i+1} = \Gamma_i$, and $\Delta_{i+1} = \Delta_i\cup
  \{w:A[h'/w]\}$.
\item If $\exfml{F}_i = \Pbb$ and $R_i = \{(h'',h'\simp h),(h'',h'\simp
  w)\}\subseteq\Gcal_i$, then $\Gcal_{i+1} = \Gcal_i\cup\{h\sim w\}$,
  $\Gamma_{i+1} = \Gamma_i$ and $\Delta_{i+1} = \Delta_i$.
\item If $\exfml{F}_i = \Cbb$ and $R_i = \{(h'',h'\simp h),(h'',w\simp
  h)\}\subseteq\Gcal_i$, then $\Gcal_{i+1} = \Gcal_i\cup\{h'\sim w\}$,
  $\Gamma_{i+1} = \Gamma_i$ and $\Delta_{i+1} = \Delta_i$.
\item If $\exfml{F}_i = \Dbb$ and $R_i = \{(h,h\simp
  h')\}\subseteq\Gcal_i$, then $\Gcal_{i+1} = \Gcal_i\cup\{h\sim
  \epsilon\}$, $\Gamma_{i+1} = \Gamma_i$, and $\Delta_{i+1} =
  \Delta_i$.
\end{itemize}
\item In all other cases, $\Gcal_{i+1} = \Gcal_i$, $\Gamma_{i+1} =
  \Gamma_i$ and $\Delta_{i+1} = \Delta_i$.
\end{itemize}

\vspace{10px}
\noindent \textbf{Lemma~\ref{lem:lim_hintikka}.}
If $F$ is a formula unprovable in $\lsfoasl$, then the limit sequent
for $F$ is a Hintikka sequent.

\begin{proof}
Let $\myseq{\Gcal^\omega}{\Gamma^\omega}{\Delta^\omega}$ be the limit
sequent.  First we show that
$\myseq{\Gcal^\omega}{\Gamma^\omega}{\Delta^\omega}$ is
finitely-consistent. Consider any
$\myseq{\Gcal}{\Gamma}{\Delta}\subseteq_f
\myseq{\Gcal^\omega}{\Gamma^\omega}{\Delta^\omega}$, we show that
$\myseq{\Gcal}{\Gamma}{\Delta}$ has no derivation. Since
$\Gcal,\Gamma,\Delta$ are finite sets, there exists $i\in \mathcal{N}$
s.t. $\Gcal\subseteq \Gcal_i$, $\Gamma\subseteq \Gamma_i$, and
$\Delta\subseteq \Delta_i$.  Moreover,
$\myseq{\Gcal_i}{\Gamma_i}{\Delta_i}$ is not provable in $\lsfoasl$,
thus $\myseq{\Gcal}{\Gamma}{\Delta}\subseteq_f
\myseq{\Gcal_i}{\Gamma_i}{\Delta_i}$ cannot be provable either.  So
condition 1, 2, and 4 in Definition~\ref{definition:hintikka_seq} hold
for the limit sequent, for otherwise we would be able to construct a
provable finite labelled sequent from the limit sequent.

In some rules (thus also in the construction of the limit sequent),
e.g., $\mand R$, a label may occur multiple times in the principal
relational atoms/formulae. In the corresponding condition for the
Hintikka sequent, we use distinct labels instead. This causes no
problems since the rule $\sim_2$ (Condition 18 of Hintikka sequent)
ensures that any formula/relational atom, in which a label is
substituted with an equivalent one, must also be in the sequent. We
explain this using $\mand R$ as an example:

Suppose $h''':A\mand B \in \Delta^\omega$ and $\{(h,h'\simp
h''),h''\sim h'''\}\subseteq\Gcal^\omega$. There must be some
$i\in\Nbb$ such that $h''':A\mand B\in\Delta_i$ and $\{(h,h'\simp
h''),h''\sim h'''\}\subseteq\Gcal_i$. There must also be some $j > i$
with $\phi(j) = (\_,\_,\approx_2,\{(h,h'\simp w)[h''/w],h''\sim
h'''\},\{\},3)$. By the first construction clause of $\approx_2$ in
Definition~\ref{definition:sequent-series}, $\Gcal_{j+1}$ contains
$(h,h'\simp w)[h'''/w]$, which is $(h,h'\simp h''')$. Now, there must
be some $k > j$ such that $\phi_k = (1,h''',A\mand B,\{h,h'\simp
h'''\},\_,\_)$. By Definition~\ref{definition:sequent-series}, either
$\Delta_{k+1} = \Delta_k\cup\{h:A\}$ or $\Delta_{k+1} =
\Delta_k\cup\{h':B\}$, whichever is not derivable. Thus Condition 8 of
Hintikka sequent is met.

With the above technique, most of other conditions in
Definition~\ref{definition:hintikka_seq} can be checked in a similar
way as the proof in~\cite{hou2013b}, here we give some cases that do
not appear in the previous work. The following cases are numbered
according to items in Definition~\ref{definition:hintikka_seq}.
\begin{enumerate}
  \setcounter{enumi}{10}
\item If $h:\exists x.A(x)\in\Gamma^\omega$, then there is some
  $i\in\Nbb$ such that $h:\exists x.A(x)\in\Gamma_i$. And there is
  some $j> i$ such that $\phi(j) = (0,h,\exists x.A(x),\_,\_,\_)$. By
  Definition~\ref{definition:sequent-series},
  $h:A(t_{j+1})\in\Gamma_{j+1}$, thus $h:A(t_{j+1})\in\Gamma^\omega$.
\item If $h:\exists x.A(x)\in\Delta^\omega$, then there is some
  $i\in\Nbb$ such that $h:\exists x.A(x)\in\Delta_i$. For each $t_n\in
  D$, there is some $j > max(i,n)$ with $\phi(j) = (1,h,\exists
  x.A(x),\_,\{h:t_n = t_n\},\_)$. By
  Definition~\ref{definition:sequent-series},
  $h:A(t_n)\in\Delta_{j+1}$, thus $h:A(t_n)\in\Delta^\omega$. 
\end{enumerate}
The other conditions can be shown similarly.
\qed
\end{proof}

\noindent \textbf{Theorem~\ref{thm:lsfoasl_complete}.}
If $F$ is valid in $\foasl$, then $F$ is provable in $\lsfoasl$.

\begin{proof}
We construct a limit sequent
$\myseq{\Gcal^\omega}{\Gamma^\omega}{\Delta^\omega}$ for $F$ following
Definition~\ref{definition:lim_seq}. Note that by the construction of
the limit sequent, we have $h_1 : F \in \Delta^\omega.$ By
Lemma~\ref{lem:lim_hintikka}, this limit sequent is a Hintikka
sequent, and therefore by Lemma~\ref{lem:Hintikka_model},
$\myseq{\Gcal^\omega}{\Gamma^\omega}{\Delta^\omega}$ is falsifiable.
This means there exists an extended model $(\Mcal,\rho)$ that
satisfies every item in $\Gcal^\omega$ and $\Gamma^\omega$ and
falsifies every item in $\Delta^\omega$, including $h_1 :
F$. Therefore $F$ is false at world $\rho(h_1)$ in that model. Thus
$F$ is not valid.
Note that the limit sequent corresponds to a Hintikka sequent with an
infinite set of closed terms. In case that the target language only
has a finite set of closed terms and constants, we can just ignore the
interpretations for the redundant ones.
\qed
\end{proof}

\section{An example derivation}
\label{app:ex_dev}

We sometimes write $r\times n$ when it is obvious that the rule $r$ is
applied $n$ times. We omit some formulae to save space. 

  \begin{center}   
    \small
\AxiomC{}
\RightLabel{\tiny $\top^* R$}
\UnaryInfC{$\myseq{\cdots
}{
\epsilon:e_3\mapsto e_4}{\epsilon:\top^*\cdots
}$}
\AxiomC{}
\RightLabel{\tiny $id$}
\UnaryInfC{$\myseq{\cdots
}{
\epsilon:e_3\mapsto e_4}{\epsilon:e_3\mapsto e_4;\cdots
}$}
\RightLabel{\tiny $\land R$}
\BinaryInfC{$\myseq{\cdots
}{
\epsilon:e_3\mapsto e_4}{\epsilon:(e_3\mapsto e_4) \land \top^*;\cdots
}$}
\AxiomC{}
\RightLabel{\tiny $\bot L$}
\UnaryInfC{$\myseq{\cdots
}{\epsilon:\bot;
\epsilon:e_3\mapsto e_4}{\cdots
}$}
\RightLabel{\tiny $\limp L$}
\BinaryInfC{$\myseq{\cdots
}{\epsilon:((e_3\mapsto e_4) \land \top^*) \limp \bot;
\epsilon:e_3\mapsto e_4}{\cdots
}$}
\RightLabel{\tiny $\forall L\times 2$}
\UnaryInfC{$\myseq{\cdots
}{\epsilon:\forall e_1,e_2.((e_1\mapsto e_2) \land \top^*) \limp \bot;
\epsilon:e_3\mapsto e_4}{\cdots
}$}
\RightLabel{\tiny $\Box L$ on Formula 1}
\UnaryInfC{$\myseq{\cdots
}{
\epsilon:e_3\mapsto e_4}{\cdots
}$}
\RightLabel{\tiny $Eq_1$}
\UnaryInfC{$\myseq{(\epsilon,h_3\simp \epsilon);\cdots
}{
h_3:e_3\mapsto e_4}{\cdots
}$}
\RightLabel{\tiny $\top^* L$}
\UnaryInfC{$\myseq{\cdots
}{
h_3:e_3\mapsto e_4;h_3:\top^*}{\cdots
}$}
\RightLabel{\tiny $\lnot R$}
\UnaryInfC{$\myseq{\cdots
}{
h_3:e_3\mapsto e_4}{h_3:\lnot\top^*;\cdots
  }$}
\noLine
\UnaryInfC{$\Pi_2$}
\DisplayProof\\[20px]

\AxiomC{}
\RightLabel{\tiny $\bot L$}
\UnaryInfC{$\myseq{\cdots
}{
h_4:\bot}{\cdots}$}
\AxiomC{$\Pi_1$}
\AxiomC{$\Pi_2$}
\RightLabel{\tiny $\mand R$}
\BinaryInfC{$\myseq{(h_3,h_1\simp h_5);\cdots
}{
h_1:e_1\mapsto e_2;h_3:e_3\mapsto e_4}{h_5:((\lnot \top^*) \mand (\lnot \top^*));\cdots}$}
\RightLabel{\tiny $\mimp L$}
\BinaryInfC{$\myseq{(h_5,h_0\simp h_4);(h_3,h_1\simp h_5)}{\cdots;h_0:((\lnot \top^*) \mand (\lnot \top^*)) \mimp \bot;h_1:(e_1\mapsto e_2);h_3:(e_3\mapsto e_4)}{\cdots}$}
\RightLabel{\tiny $E$}
\UnaryInfC{$\myseq{(h_0,h_5\simp h_4);(h_3,h_1\simp h_5)}{\cdots;h_0:((\lnot \top^*) \mand (\lnot \top^*)) \mimp \bot;h_1:(e_1\mapsto e_2);h_3:(e_3\mapsto e_4)}{\cdots}$}
\RightLabel{\tiny $A$}
\UnaryInfC{$\myseq{(h_0,h_1\simp h_2);(h_2,h_3\simp h_4)}{\cdots;h_0:((\lnot \top^*) \mand (\lnot \top^*)) \mimp \bot;h_1:(e_1\mapsto e_2);h_3:(e_3\mapsto e_4)}{\cdots}$}
\RightLabel{\tiny $E\times 2$}
\UnaryInfC{$\myseq{(h_1,h_0\simp h_2);(h_3,h_2\simp h_4)}{\cdots;h_0:((\lnot \top^*) \mand (\lnot \top^*)) \mimp \bot;h_1:(e_1\mapsto e_2);h_3:(e_3\mapsto e_4)}{h_4:\bot;h_2:\bot}$}
\RightLabel{\tiny $\mimp$ R}
\UnaryInfC{$\myseq{(h_1,h_0\simp h_2)}{\cdots;h_0:((\lnot \top^*) \mand (\lnot \top^*)) \mimp \bot;h_1:(e_1\mapsto e_2)}{h_2:(e_3\mapsto e_4) \mimp \bot;h_2:\bot}$}
\RightLabel{\tiny $\lnot L$}
\UnaryInfC{$\myseq{(h_1,h_0\simp h_2)}{\cdots;h_2:\lnot ((e_3\mapsto e_4) \mimp \bot);h_0:((\lnot \top^*) \mand (\lnot \top^*)) \mimp \bot;h_1:(e_1\mapsto e_2)}{h_2:\bot}$}
\RightLabel{\tiny $\forall L$}
\UnaryInfC{$\myseq{(h_1,h_0\simp h_2)}{\cdots;h_2:\forall e_2. \lnot ((e_3\mapsto e_2) \mimp \bot);h_0:((\lnot \top^*) \mand (\lnot \top^*)) \mimp \bot;h_1:(e_1\mapsto e_2)}{h_2:\bot}$}
\RightLabel{\tiny $\exists L$}
\UnaryInfC{$\myseq{(h_1,h_0\simp h_2)}{\cdots;h_2:\exists e_1 \forall e_2. \lnot ((e_1\mapsto e_2) \mimp \bot);h_0:((\lnot \top^*) \mand (\lnot \top^*)) \mimp \bot;h_1:(e_1\mapsto e_2)}{h_2:\bot}$}
\RightLabel{\tiny $\Box L$ on Formula 5}
\UnaryInfC{$\myseq{(h_1,h_0\simp h_2)}{\cdots;h_0:((\lnot \top^*) \mand (\lnot \top^*)) \mimp \bot;h_1:(e_1\mapsto e_2)}{h_2:\bot}$}
\RightLabel{\tiny $\mimp R$}
\UnaryInfC{$\myseq{\cdots}{h_0:((\lnot \top^*) \mand (\lnot \top^*)) \mimp \bot}{h_0:(e_1\mapsto e_2) \mimp \bot}$}
\RightLabel{\tiny $\lnot L$}
\UnaryInfC{$\myseq{\cdots}{h_0:\lnot ((e_1\mapsto e_2) \mimp \bot);h_0:((\lnot \top^*) \mand (\lnot \top^*)) \mimp \bot}{}$}
\RightLabel{\tiny $\forall L$}
\UnaryInfC{$\myseq{\cdots}{h_0:\forall e_2. \lnot ((e_1\mapsto e_2) \mimp \bot);h_0:((\lnot \top^*) \mand (\lnot \top^*)) \mimp \bot}{}$}
\RightLabel{\tiny $\exists L$}
\UnaryInfC{$\myseq{\cdots}{h_0:\exists e_1 \forall e_2. \lnot ((e_1\mapsto e_2) \mimp \bot);h_0:((\lnot \top^*) \mand (\lnot \top^*)) \mimp \bot}{}$}
\RightLabel{\tiny $\Box L$ on Formula 5}
\UnaryInfC{$\myseq{}{h_0:\Acal;h_0:((\lnot \top^*) \mand (\lnot \top^*)) \mimp \bot}{}$}
\RightLabel{\tiny $\lnot R$}
\UnaryInfC{$\myseq{}{h_0:\Acal}{h_0:\lnot (((\lnot \top^*) \mand (\lnot \top^*)) \mimp \bot)}$}
\RightLabel{\tiny $\limp R$}
\UnaryInfC{$\myseq{}{}{h_0:\Acal\limp \lnot (((\lnot \top^*) \mand (\lnot \top^*)) \mimp \bot)}$}
\DisplayProof\\[10px]

N.B. The sub-derivation $\Pi_1$ is similar to $\Pi_2$.
\end{center}

\end{document}